\documentclass[twocolumn,prb,showpacs,preprintnumbers,amsmath,superscriptaddress,amssymb]{revtex4}

\usepackage{graphicx}
\usepackage{dcolumn}
\usepackage{bm}

\begin{document}

\title{Efficient real frequency solver for dynamical mean field theory}

\author{Y. Lu}
  \affiliation{Max Planck Institute for Solid State Research, Heisenbergstra{\ss}e 1, 70569 Stuttgart, Germany}
  \affiliation{Department of Physics and Astronomy, University of British Columbia, Vancouver, British Columbia V6T1Z1, Canada}
  \affiliation{Max Planck Institute for Chemical Physics of Solids, N{\"o}thnitzerstra{\ss}e 40, 01187 Dresden, Germany}
\author{M. H{\"o}ppner}
  \affiliation{Max Planck Institute for Solid State Research, Heisenbergstra{\ss}e 1, 70569 Stuttgart, Germany}
\author{O. Gunnarsson}
  \affiliation{Max Planck Institute for Solid State Research, Heisenbergstra{\ss}e 1, 70569 Stuttgart, Germany}
\author{M. W. Haverkort}
  \affiliation{Max Planck Institute for Solid State Research, Heisenbergstra{\ss}e 1, 70569 Stuttgart, Germany}
  \affiliation{Department of Physics and Astronomy, University of British Columbia, Vancouver, British Columbia V6T1Z1, Canada}
  \affiliation{Max Planck Institute for Chemical Physics of Solids, N{\"o}thnitzerstra{\ss}e 40, 01187 Dresden, Germany}

\date{\today}

\begin{abstract}
We here present how a self-consistent solution of the dynamical mean field theory equations can be obtained using exact diagonalization of an Anderson impurity model with accuracies comparable to those found using renormalization group or quantum Monte Carlo methods. We show how one can solve a correlated quantum impurity coupled to several hundred uncorrelated bath sites, using a restricted active basis set. The number of bath sites determines the resolution of the obtained spectral function, which consists of peaks with an approximate spacing proportional to the band width divided by the number of bath sites. The self-consistency cycle is performed on the real frequency axis and expressed as numerical stable matrix operations. The same impurity solver has been used on ligand field and finite size cluster calculations and is capable of treating involved Hamiltonians including the full rotational invariant Coulomb interaction, spin-orbit coupling, and low-symmetry crystal fields. The proposed method allows for the calculation of a variety of correlation functions at little extra cost.
\end{abstract}


\maketitle

\section{Introduction}

Theoretical understanding of correlated electron systems is often hindered by the exponential scaling of the computation time and memory required as a function of system size. For systems where the local density or Hartree Fock approximations fail, there exists a real computational problem. Obtaining quantum chemical \textit{ab initio} solutions is impossible for many systems.\cite{Kohn:1999zza} Even small systems containing only three or four open $d$- or $f$-shell ions can be too large to compute. Nonetheless, one can obtain information on open shell compounds in the approximation of a single correlated site interacting with mean-field approximated neighbors or bath sites. Such an embedded impurity in a mean-field approximated bath can either be realized by the requirement that the density or the one particle Green's function is equivalent on the mean-field approximated sites and the impurity. The latter results in the dynamical mean field theory (DMFT).\cite{Metzner:1989eq, MullerHartmann:1989uj, Georges:1992to, Jarrell:1992vo, Rozenberg:1992wv, Pruschke:1995ho, Georges:1996un, Kotliar:2004ua, Maier:2005tj, Kotliar:2006fl, Held:2007fi, Vollhardt:2012tl, Lechermann:2006iq} In either case the one-particle energies and hopping integrals can be obtained directly from density functional theory,\cite{Gunnarsson:1989tf, Anisimov:1991vf, Held:2001cv,Aryasetiawan:2004kq, Lechermann:2006iq, Haverkort:2012du} or Hartree Fock calculations.\cite{Hozoi:2007gl}

In the case of transition metal oxides, the mean-field approximated neighbors are, in first approximation, the ligand O atoms. If one only includes a single transition metal impurity interacting with ligand orbitals, one obtains  multiplet ligand field theory.\cite{Griffith:1957bs, Atkins:2010uh, Haverkort:2012du} Ligand field theory is one of the oldest methods used to solve the Schr{\"o}dinger equation. Nonetheless, for correlated insulators it is still a very powerful approximation. For correlated metals, ligand field theory is clearly not sufficient. In this case one needs to include a full band, which leads to an Anderson impurity model. In an (cluster) Anderson impurity model there are $N_{\tau}$ partially filled impurity levels (spin, orbital and cluster site) with correlations between the electrons occupying these levels, each interacting with $N_b$ partially filled bath sites. This is a highly nontrivial problem whereby in general the basis size scales exponentially in the number of total sites and levels ($N_{\tau}+N_{\tau} \times N_b$) included in the problem. Nonetheless, an infinite Anderson impurity model can be solved. Several methods are available; each has its virtues, but all have shortcomings.

Since the introduction of DMFT there has been an enormous development on how to solve an Anderson impurity Hamiltonian. For the single-site Hubbard model there exist beautiful solutions using numerical renormalization group (NRG) theory\cite{Bulla:1999js,  Bulla:2001uj, Bulla:2005en, Bulla:2006ek, Bulla:2008jn, Pruschke:2000eq, Byczuk:2007en, Bauer:2009kw, Bauer:2009hv, Zitko:2009bv} or density renormalization group theory.\cite{Garcia:2004cs, Nishimoto:2004fa, Hallberg:2006jg, Miranda:2008cs} These methods are hard to apply to situations with multiple interacting orbitals or sites. Hirsch Fye (HF),\cite{Hirsch:1986ud, Jarrell:1992vo, Rozenberg:1992wv, Georges:1992ur, Ulmke:1995vz, Blumer:2007it} and continuous time (CT) \cite{Gull:2011jd, Gull:2011hw, Rubtsov:2004wo, Rubtsov:2005iw, Wernet:2006ko, Wernet:2006iz, Gull:2008cm} quantum Monte Carlo (QMC) methods can be rather efficient for the single site Hubbard model as well as some extensions including several correlated fermions, but seem to have problems with low symmetry interactions and systems where the Green's function has off-diagonal terms. A further drawback of QMC implementations is the use of imaginary instead of real frequencies, which leads to an ill-conditioned inversion problem.\cite{Jarrell:1996uo, Beach:2000um, Gunnarsson:2010hw}

Exact diagonalization (ED) techniques\cite{Caffarel:1994vs, Jaklic:1994ii, Sangiovanni:2006ez, Capone:2007hd, Koch:2008bx, Senechal:2010ga, Liebsch:2011cd, Zgid:2011do, Weber:2012fc, Go:2013tl, Zgid:2012ck, Lin:2013en} can be applied very generally, are implemented using real frequencies, and pose no requirement on the Hamiltonian other than that it should be reasonably sparse. The problem with this method, though, is that the mean-field approximated bath has to be represented by a small number of discrete states in order to keep the exponentially growing many-body Hilbert space tractable.\cite{Potthoff:2001ft} This can be improved by selecting a certain subset of many-body states as the basis. Reasonable results for a single Ce $4f$ shell interacting with a free-electron-like band have been obtained by selecting only certain basis functions.\cite{Gunnarsson:1983vb, Gunnarsson:1983uf, Gunnarsson:1985vp} The question of which states to include can be formalized using a configuration interaction\cite{Sherrill:1999tg, VanOosten:1990tj, Tanaka:1994vx, Zgid:2011do, Lin:2013en} or coupled cluster expansion.\cite{Nakatsuji:1977ti, Monkhorst:1977vv, Jeziorski:1981vo, Lindgren:1987ui} For DMFT on the Bethe lattice one can, with the use of a configuration interaction expansion of the basis, optimize the basis in such a way that one can obtain a converged ground state.\cite{Zgid:2011do, Go:2013tl, Zgid:2012ck, Lin:2013en} The configurations included in these calculations are optimized to represent the ground state, but not the excited states needed in the calculation of the one-particle Green's function. Presently, configuration interaction calculations do not converge the Green's function, which is an important ingredient in DMFT.

Here we show how a general solution of the dynamical mean-field equations can be obtained. We use an ED technique which can include the full rotational invariant Coulomb interaction, spin-orbit coupling as well as low-symmetry interactions. In the current paper we show the solution of the Hubbard model on a Bethe lattice at $T=0$. Extensions to higher temperatures might be possible, but have not been tested.\cite{Jaklic:1994ii, Prelovsek:2013uw} The impurity solver has been used in several multiorbital or multisite calculations\cite{Haverkort:2012du, Glawion:2012gs, Glawion:2011jf} and the inclusion of five (open $d$-shell) or seven (open $f$-shell) correlated orbitals or eight (two dimensional cluster) correlated sites coupled to several hundreds of uncorrelated bath sites is in principle possible, albeit not yet implemented in the DMFT scheme. 

The method presented here is similar to a recently independently implemented variational approach based on the configuration interaction expansion by Lin and Demkov.\cite{Lin:2013en} As shown in their publication, it is crucial to use an optimized bath parametrization, which they obtain with the use of natural orbitals. The main difference with our method is that we do not use a configuration interaction expansion of the many-body basis states, but search for the $\approx 10^9$ Slater determinants with the largest contribution in the full basis. We thus do not need to set the configurations before the calculation starts, but establish during the calculation which determinants need to be included. This leads to a different basis for the ground state and excited states. The resulting method allows for the inclusion of several hundreds of discretized bath sites. On this basis we are able to find a converged ground state as well as a converged Green's function.

In the main part of the paper we first introduce how to implement the DMFT self-consistency loop using numerically stable matrix operations on real frequency representations of the Green's functions and self-energy. We continue by showing how one can solve the Anderson impurity problem using ED including several hundred bath sites. The paper is written to convey the general idea and overview of the method without too much detail. Additional details and mathematically rigorous definitions are placed in the appendixes. After the method is introduced, we show results for the Hubbard model on a Bethe lattice as a function of $U$ and number of discretized bath sites. An important result is that the critical value of $U$, for which the metal insulator transition takes place, depends on the number of sites included. We compare our results to analytically known sum rules, to NRG results by Bulla,\cite{Bulla:1999js,  Bulla:2001uj, Bulla:2005en, Bulla:2006ek, Bulla:2008jn} as well as to results obtained from QMC calculations. Good agreement in terms of the quasiparticle weight and bandwidth is obtained. The same is true for the Hubbard bands, which show the same weight, position, and width as found in other methods.

Appendix \ref{app:notation} discusses the notation used in this paper. In Appendix \ref{app:Gtrans} we show the relation between different representations of the Green's function. In Appendix \ref{app:lanc} the Lanczos algorithm is explained. In Appendix \ref{app:lancsparse} we discuss the optimized many electron sparse Slater determinant basis used in the Lanczos algorithm. Appendix \ref{app:natorb} explains the optimized one particle basis functions or orbitals used. In Appendix \ref{app:polered} we discuss the reduction of poles in the Green's functions used, which is equivalent to choosing an optimized number of one-electron basis functions to represent the Anderson impurity Hamiltonian.

\section{The DMFT self-consistency loop}
\label{section:loop}

 \begin{figure}
    \includegraphics[width=0.45\textwidth]{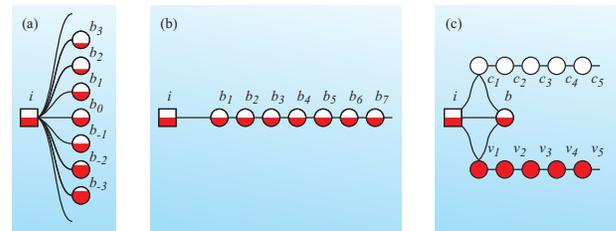}
    \caption{(Color online) Possible bath geometries. The impurity is labeled by $i$ and represented by a square. The bath sites are labeled by $b_{i}$ [panels (a) and (b)] or by $b$, $c_i$, and $v_i$. The site occupation is indicated by the filling. For efficient calculations bath sites should either be occupied or empty.}
    \label{Fig:Structure}
 \end{figure}

The self-consistency loop in the DMFT calculations breaks down in four parts.\cite{Koch:2008bx} In the calculation one repeats steps 1 to 4 until the bath and impurity Green's function are converged and do not change between loops. Most DMFT implementations use Green's functions represented on imaginary frequencies in the self-consistent loop. As the transformation between Green's functions represented on the real and imaginary axis is bijective, this is possible without loss of information. The disadvantage is that the transformation is also ill conditioned, which requires one to use extraordinarily large numerical accuracy on the imaginary axis.\cite{Beach:2000um} To circumvent these numerical problems we perform the entire calculation using Green's functions represented on the real-frequency axis.

The Green's functions and self-energy are expressed as a sum over delta functions, which can be related to the resolvent of a matrix. One can apply unitary transformations on the matrix representing the Green's function without changing the Green's function and this is used during the self-consistency loop. Important to note is that during the entire calculation the Green's function is expressed as a discrete sum of delta functions with zero width with variable weight and energy. These poles representing the Green's functions and self-energy are never replaced by a sum over Lorentzians with finite broadening inside the self-consistency loop. A finite broadening is included only when plots on the real axis are made. More details about the different representations can be found in Appendix \ref{app:Gtrans}. Below we show the DMFT self-consistency loop as implemented in this paper based on Green's functions and self energies represented by a sum of delta functions on the real-frequency axis.

\subsubsection{From bath Green's function to Anderson impurity Hamiltonian}

We start our self-consistency loop with a known ($T=0$) retarded bath Green's function [$G_b(\omega)$]. This could be the noninteracting Green's function [$G_0(\omega)$] if no better approximation is known. The first task is to define the Anderson impurity Hamiltonian ($H_A$), given the bath Green's function. This is a straight-forward task. The bath Green's function is defined and stored by $N_b+1$ numerical values of $\alpha_i$ and $N_b$ values of $\beta_i$ as
\begin{equation}
\label{eq:Gb}
G_b(\omega)=\frac{1}{\omega-\alpha_1^b-\sum_{j=1}^{N_b} \frac{{\beta_j^b}^2}{\omega-\alpha_{j+1}^b}}.
\end{equation}
$N_b$ defines the number of discretization points of the Green's function as well as the number of bath orbitals in the Anderson impurity Hamiltonian. This form of the Green's function can easily be obtained from any other representation as shown in Appendix \ref{app:Gtrans}. For cases where the impurity consists of multiple orbitals, sites or spin states, $\alpha$ and $\beta$ are matrices of dimension $N_{\tau}$ by $N_{\tau}$, with $\tau$ labeling the internal spin, orbital and site degree of freedom of the impurity. We as much as possible suppress summations over $\tau$ using the definitions as given in Appendix \ref{app:notation}. 

The Anderson impurity Hamiltonian has, besides the additional correlations on the impurity site, an interaction of $\beta_j$ with a bath site at onsite energy $\alpha_{j+1}$. Graphically, one can represent this Hamiltonian and Green's function with a single impurity interacting with $N_b$ bath sites as shown in panel (a) of figure \ref{Fig:Structure}. In formula this is
\begin{align}
\nonumber
H_A&=H_i + \alpha_1^b a^{\dag}_i a^{\phantom{\dag}}_i \\&+ \sum_{j=1}^{N_b} \beta_j^b \left(a^{\dag}_i a^{\phantom{\dag}}_{b_j}+ a^{\dag}_{b_j}a^{\phantom{\dag}}_i\right) + \alpha_{j+1}^b a^{\dag}_{b_j} a^{\phantom{\dag}}_{b_j},
\end{align}
with $i$ ($b_j$) labeling the impurity (bath) and $j$ an index for the different discretized bath states. $H_i$ is the many-body Hamiltonian which only acts on the impurity sites,
\begin{align}
\nonumber
H_i &= \sum_{\tau,\tau'} \epsilon_{\tau,\tau'} a^{\dag}_{i,\tau} a^{\phantom{\dag}}_{{i,\tau'}} \\&+ \sum_{\tau,\tau',\tau'',\tau'''} U_{\tau,\tau',\tau'',\tau'''} a^{\dag}_{i,\tau} a^{\dag}_{i,\tau'} a^{\phantom{\dag}}_{i,\tau''}a^{\phantom{\dag}}_{i,\tau'''},
\end{align}
with $\epsilon$ and $U$ numerical parameters defining the one- and two-electron parts of the many-body Hamiltonian and $\tau$ being an index for the different fermion quantum states (orbital, spin, and site) within the impurity, which here has been written out explicitly.

\subsubsection{From Anderson impurity Hamiltonian to impurity Green's function}

Once the Anderson impurity Hamiltonian is known, the ground state of this Hamiltonian is obtained and the impurity Green's function [$G_c(\omega)$] is calculated. This step is discussed in more detail in the next section. Here we just state that the resulting impurity Green's function can be expressed as
\begin{equation}
G_c(\omega)=\frac{1}{\omega-\alpha_1^c-\sum_{j=1}^{N_c} \frac{{\beta_j^c}^2}{\omega-\alpha_{j+1}^c}},
\end{equation}
with $\alpha_i^c$ and $\beta_i^c$ numerical values defining the Green's function. $N_c$ defines the number of poles in the impurity Green's function and should be at least as large as the number of poles in the bath Green's function and probably even slightly larger. In the current paper we use $N_c=1000$. 

\subsubsection{From impurity and bath Green's function to impurity self-energy}

From the bath Green's function and the impurity Green's function one can calculate the impurity self-energy:\cite{Metzner:1989eq, MullerHartmann:1989uj, Georges:1992to, Jarrell:1992vo, Rozenberg:1992wv, Pruschke:1995ho, Kotliar:2004ua, Georges:1996un, Maier:2005tj, Kotliar:2006fl, Held:2007fi, Vollhardt:2012tl, Lechermann:2006iq}
\begin{equation}
\Sigma_c(\omega)=G_b(\omega)^{-1}-G_c(\omega)^{-1}.
\end{equation}
Using the previous definitions of $G_b(\omega)$ and $G_c(\omega)$ this yields
\begin{equation}
\Sigma_c(\omega) = \alpha_1^c-\alpha_1^b+\sum_{j=1}^{N_c} \frac{{\beta_j^c}^2}{\omega-\alpha_{j+1}^c}-\sum_{j=1}^{N_b} \frac{{\beta_j^b}^2}{\omega-\alpha_{j+1}^b},
\end{equation}
which can be regrouped as
\begin{equation}
\label{eq:sigma}
\Sigma_c(\omega) = \alpha_1^{\Sigma} + \sum_{j=1}^{N_{\Sigma}} \frac{{\beta_j^{\Sigma}}^2}{\omega-\alpha_{j+1}^{\Sigma}},
\end{equation}
with $\alpha^{\Sigma}$ and $\beta^{\Sigma}$ numerical values defining the self-energy as a function of $\omega$.

In order for the self-energy to represent a physical quantity, ${\beta_j^{\Sigma}}^2$ must be larger than zero. This is fulfilled if for any pole at energy $\alpha_j^b$ originating from the bath Green's function there is a pole originating from the impurity Green’s function at the same energy ($\alpha_{j'}^c=\alpha_j^b$) with a larger weight (${\beta_{j'}^c}^2-{\beta_{j}^c}^2>0$). For calculations with infinity precision math and $N_c\to\infty$, this is the case and the self-energy will be physical. In real calculations  with $N_b$ of the order of several hundred and with computers with 16 digits accuracy, this will not be the case. The self-energy can be made physical by merging poles with a negative weight with poles in the neighborhood. If one orders the poles representing the self-energy in equation (\ref{eq:sigma}) such that $\alpha_j^{\Sigma} < \alpha_{j+1}^{\Sigma}$, then a pole with index $j$ and ${\beta_j^\Sigma}^2<0$ is merged with the poles $j-1$ and $j+1$. The weight ($\beta^2$) and energy ($\alpha$) of the new poles is chosen such to conserve locally the zeroth and first moment and only introduce small errors in the higher moments. The removal of the pole with the smallest weight is done first and this is repeated until all poles have a positive weight. This procedure reduces the number of poles in the self-energy ($N_{\Sigma}$) to a maximum of the number of poles in the impurity Green's function ($N_c$). The number can be smaller if after merging a pole with negative weight with a neighbor pole the result is still negative. Details of this procedure are presented in Appendix \ref{app:polered}.

\subsubsection{From impurity self-energy and noninteracting Green's function to the new bath Green's function}

The new bath Green's function can be calculated by the noninteracting Green's function $G_0(\omega)$ and the impurity self-energy $\Sigma_c(\omega)$. We take
\begin{equation}
\label{eq:G0}
G_0(\omega)=\frac{1}{\omega-\alpha_1-\sum_{j=1}^{N_0} \frac{\beta_j^2}{\omega-\alpha_{j+1}}},
\end{equation}
with $\alpha_j$ and $\beta_j$ numerical values defining the noninteracting Green's function which is represented by $N_0+1$ discrete poles as the resolvent of an Anderson impurity matrix. The relation between the representation given here and the density of states as one would obtain in a DFT calculation is given in Appendix \ref{app:Gtrans}. The new bath Green's function can be obtained from the self-energy and the noninteracting Green's function:
\begin{align}
\label{eq:bathnew}
\nonumber
G_b^{new}(\omega)&=\frac{1}{\omega-\alpha_1-\sum_{j=1}^{N_0} \frac{\beta_j^2}{\omega-\alpha_{j+1}-\Sigma_c(\omega)}}\\
&=\frac{1}{\omega-\alpha_1-\sum_{j=1}^{N_0} \frac{\beta_j^2}{\omega-\alpha_{j+1}-\alpha_1^{\Sigma} - \sum_{j'=1}^{N_{\Sigma}} \frac{{\beta_{j'}^{\Sigma}}^2}{\omega-\alpha_{j'+1}^{\Sigma}}}}.
\end{align}
The sum over $j$ and $j'$ can be simplified and combined into a single sum by the diagonalization of the Anderson impurity matrix ($N_0$ times, for $j=1$ to $j=N_0$):
\begin{align}
\label{eq:diagonalization}
\nonumber &\sum_{j'=(j-1)(N_{\Sigma}+1)+1}^{j(N_{\Sigma}+1)}\frac{{\beta_{j'}^{b}}^2}{\omega-\alpha_{j'+1}^b}\\
=&\frac{\beta_j^2}{\omega-\alpha_{j+1}-\alpha_1^{\Sigma} - \sum_{j'=1}^{N_{\Sigma}} \frac{{\beta_{j'}^{\Sigma}}^2}{\omega-\alpha_{j'+1}^{\Sigma}}}.
\end{align}
The resulting bath Green's function is:
\begin{equation}
G_b^{new}(\omega)=\frac{1}{\omega-\alpha_1^b-\sum_{j=1}^{N_b} \frac{{\beta_j^b}^2}{\omega-\alpha_{j+1}^b}},
\end{equation}
with $\alpha_1^b=\alpha_1$ and $\beta_j^b$ and $\alpha_j^b$ numerical values obtained from equation (\ref{eq:diagonalization}).
The number of poles in the new bath Green's function ($N_b$) is equal to $N_0 \times (N_{\Sigma}+1)$, which can become so large that it is problematic in further calculations. The reduction of the number of poles in $G_b^{new}$ is discussed in Appendix \ref{app:polered}.

The calculation of the new bath Green's function, by adding the self-energy to the noninteracting Green's function as presented in equation (\ref{eq:bathnew}), feels slightly different from the algorithm presented in most papers. \cite{Koch:2008bx,Georges:1996un} The current algorithm does not require the explicit calculation of a local Green's function. The simplification of equation (\ref{eq:bathnew}) furthermore only requires matrix diagonalization, a standard and numerical stable algorithm. During the entire self-consistency loop the Green's functions are defined as a discrete sum of poles with zero broadening. The resulting bath Green's function is given by a set of poles whose energy and weight can be different from the starting bath Green's function. Nevertheless, $G_b^{new}(\omega)$ has the same form as $G_b(\omega)$ from which the first step of the DMFT self-consistency loop started. After the calculation of $G_b^{new}$ one can restart the loop until convergence is reached.

\section{Impurity solver}

The dynamical mean field self-consistency loop requires one to solve an Anderson impurity model. The Anderson impurity Hamiltonian can be represented as a matrix. The ground state ($\psi_0$) is given as the eigenfunction of this matrix with the lowest eigenenergy. Once the ground state has been calculated, the Green's function is defined as
\begin{equation}
G(\omega) = g^+(\omega)-g^-(-\omega)^*,
\end{equation}
with
\begin{equation}
g^+(\omega)=\lim_{\Gamma\to0^+} \left\langle \psi_0 \left| a^{\phantom{\dag}}_i \frac{1}{\omega-H_A+\mathrm{i}\frac{\Gamma}{2}} a^{\dag}_i \right| \psi_0 \right\rangle,
\end{equation}
and
\begin{equation}
g^-(\omega)=\lim_{\Gamma\to0^+} \left\langle \psi_0 \left| a^{\dag}_i \frac{1}{\omega-H_A+\mathrm{i}\frac{\Gamma}{2}} a^{\phantom{\dag}}_i \right| \psi_0 \right\rangle.
\end{equation}
Here $a^{\dag}_i$ ($a^{\phantom{\dag}}_i$) creates (annihilates) an electron at the impurity site.

The definition of the Green's function requires one to calculate (twice) the resolvent of the Hamiltonian, which, in general, is a computationally involved task. For the special case where the Hamiltonian is tridiagonal, with $\varphi_0=a^{\dag}_i | \psi_0 \rangle$ ($\varphi_0=a^{\phantom{\dag}}_i | \psi_0 \rangle$) the first element of the matrix, calculating its resolvent is trivial and can be written as a continued fraction:
\begin{align}
\label{eq:contfracmain}
\nonumber
\left(
\begin{array}{ccccc}
\omega-a_1    &-b_1    & 0      & 0      & 0        \\
-b_1   &\omega-a_2    &-b_2    & 0      & 0        \\
0      &-b_2    & \ddots & \ddots & 0        \\
0      & 0      & \ddots & \ddots &-b_{n}    \\
0      & 0      & 0      &-b_{n}  &\omega-a_{n+1} 
\end{array}
\right)^{-1}_{[1,1]}\\
=\frac{1}{\omega-a_1-\frac{b_1^2}{\omega-a_2-\frac{b_2^2}{\omega-\hdots}}}.
\end{align} 
Creating the Hamiltonian in tridiagonal form is done using a Lanczos algorithm which creates the Krylov basis as:
\begin{equation}
\varphi_n = H^n a^{\dag}_i | \psi_0 \rangle.
\end{equation}
After orthonormalization, the Hamiltonian is tridiagonal on this basis and the Green's function can be obtained using equation (\ref{eq:contfracmain}).

Although the Lanczos algorithm works great on large sparse matrices, the problem encountered for an impurity coupled to a partially filled band has not been generally solved. The reason is the exponentially fast growing number of basis states needed. If one works on a basis of single Slater determinants, then the number of Slater determinants needed for a half-filled band approximated by 300 poles is $(300! / (150!)^2)^2 \approx 8.8 \times 10^{177}$. Storing a single vector of this format is far beyond reach of any computational method. Luckily, one can reduce the number to far below $10^9$, which can be handled with current computers. This can be done because not all of the $10^{177}$ determinants are equally important. The state where in a solid all electrons sit in one corner of the crystal and the rest of the crystal has no electrons is so high in energy and so unlikely, that one can safely neglect it in the calculation. The method used here searches for the $10^9$ most important determinants in the total space available and uses these to represent the ground state.

The amount of optimization possible depends highly on the Hamiltonian as well as on the one-particle orbitals used to create the Slater determinants. Optimization works generally better when the Hamiltonian spreads over a larger energy scale, with more or less empty and occupied orbitals. Although this is not something one can choose, nature often provides one with a separation of energy scales. Most solids have a separation in bands according to their atomic orbital character. The different character of bands can be used and for real calculations optimizing the one particle orbitals can mean the difference between a trivial and an impossible calculation. 

The importance of the optimization of the one-electron orbitals used in the calculation becomes clear if one looks at a noninteracting system. For the case of noninteracting electrons, one can easily write down the ground state as a single Slater determinant, which is a product of all Bloch waves with energies smaller than the Fermi energy. If one would not choose the Bloch waves as the one particle basis, but some local orbital basis, then each orbital can be partially occupied and an exponential growing number of Slater determinants is needed as a function of system size. 

For correlated systems, the one-electron basis that leads to a ground state that can be represented by a minimal number of Slater determinants, is a basis based on natural orbitals. This is a one particle basis set defined such that the density matrix for the ground state of the many-body Hamiltonian is diagonal. The disadvantage of such a basis set is that one can only obtain it after the ground state calculation is finished. As all our calculations are done iteratively, this is not a real problem and an optimal basis set is determined together with the ground state.

For fully correlated systems we do not know better single Slater determinant basis sets than the natural orbital basis set. For impurity models, where only a few orbitals have full correlations and the others are treated on a (dynamical) mean field level, the introduction of natural orbitals mixes correlated and mean-field approximated sites. This is not convenient as it complicates the Hamiltonian and results in a fully correlated problem. We therefore only allow basis rotations within the correlated orbital set and within the mean-field approximated orbital set, but do not mix these two different orbital sets.

In order to realize an optimized basis without mixing correlated and uncorrelated fermions, we need to define a way to rotate the one particle basis of the bath and impurity such that a minimum number of Slater determinants is needed in the full many-body calculation without mixing the bath orbitals with the impurity orbitals. In figure \ref{Fig:Structure} we show three different possible representations of the impurity problem, which are related to each other by a unitary transformation of the bath orbitals.

 \begin{figure*}
    \includegraphics[width=0.95\textwidth]{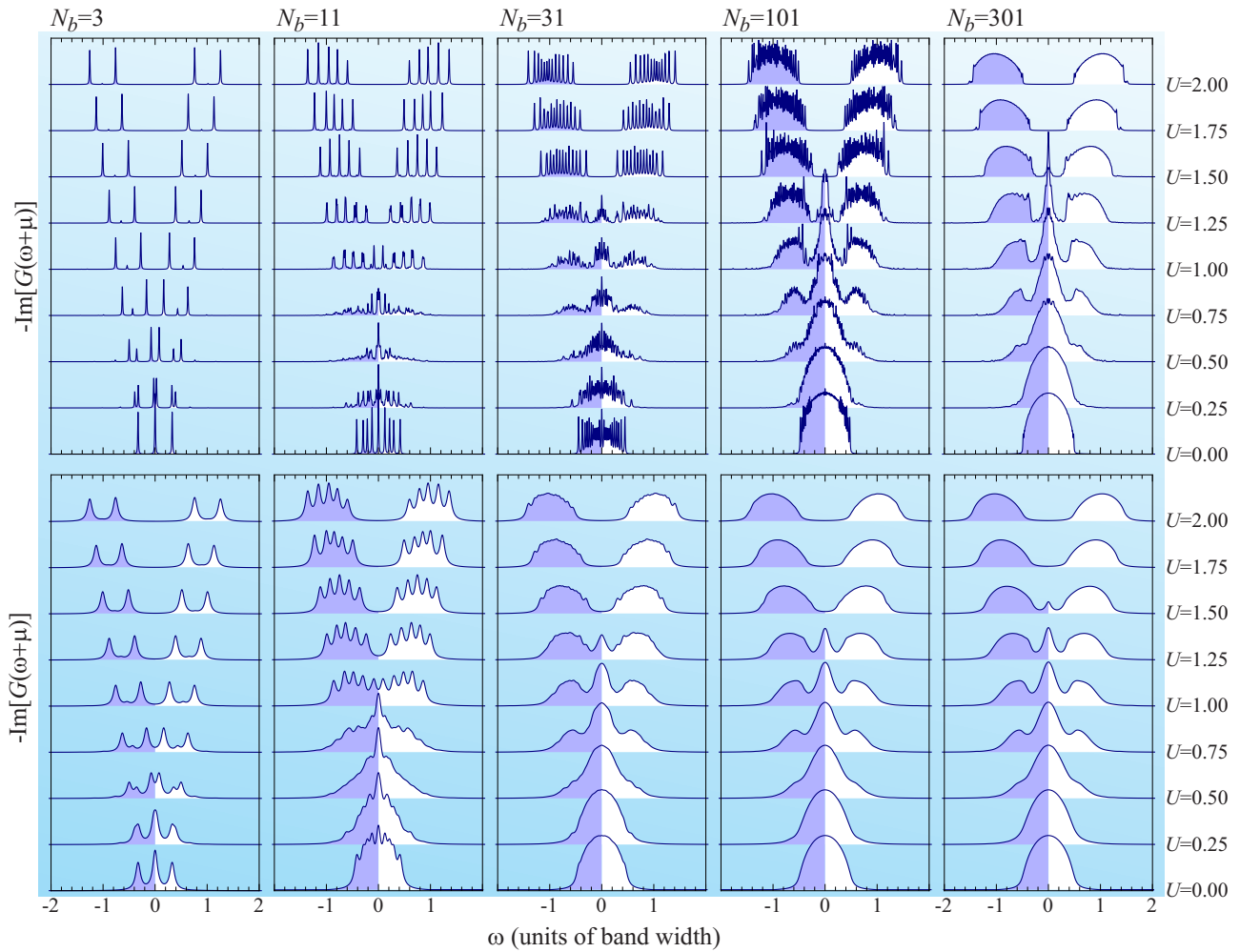}
    \caption{(Color online) All panels show the DMFT impurity Green's function for different values of $U$ ranging from 0 to 2 in steps of 0.25 in units of the band-width of $G_0$. The different columns show the spectral function for 3, 11, 31, 101 or 301 poles in the bath Green's function and thus sites in the Anderson impurity calculation. The impurity Green's function in all calculations consists of 1000 poles. The panels in the top row show the spectral functions broadened with a Lorentzian of full width at half maximum 0.01. The panels in the bottom row show the same spectral functions as the top row, but broadened with a Lorentzian of full width at half maximum 0.1.}
    \label{Fig:Results}
 \end{figure*}

The representation as shown in figure \ref{Fig:Structure} (a) has the advantage that bath orbitals with a high onsite energy are basically empty and bath orbitals with a low energy are basically fully occupied. The disadvantage is that each bath orbital directly interacts with the impurity site and therefore is important. One can make a unitary transformation of the bath sites and change the bath geometry such that the impurity site only interacts with one bath orbital, which again interacts with one other bath orbital etc. as shown in figure  \ref{Fig:Structure} (b). In this geometry the bath orbitals further away from the impurity are less important than those close to the impurity. Each bath orbital is partially occupied and the ground state is given by an exponential growing number of Slater determinants when the number of bath sites is increased. The solution is to couple the impurity to two separate chains, one representing the occupied states of the bath and one representing the unoccupied states of the bath. In order to be able to choose any filling of the impurity and still only have fully occupied or fully empty states, one needs an additional bath site, which for an impurity with a filling of $n$ has a filling of $1-n$. The resulting total number of electrons is always integer. This bath geometry is shown in figure \ref{Fig:Structure} (c). 

Within our calculations we obtain a similar geometry as shown in figure \ref{Fig:Structure} (c) automatically. We require the density matrix of the impurity as well as the density matrix of the bath to be diagonal. In order to reach this situation, we need a starting point, which allows one to calculate the ground state and density matrix of a basis including hundreds of orbitals. We therefore define a noninteracting reference system which gives a good starting point. Using this reference basis leads exactly to the bath geometry as shown in figure \ref{Fig:Structure} (c). In Appendix \ref{app:natorb} we discuss the transition between the different representations in more detail.

\section{Results}

\subsection{Dependence on $U$ and number of bath sites}

In order to test the algorithm as described in the previous two sections, we calculate the Hubbard model on a Bethe lattice for different values of the Coulomb interaction $U$. The obtained impurity Green's function can be seen in figure \ref{Fig:Results}. The impurity Green's functions are represented by a sum of $N_c$ delta functions at some energy and with some weight, such that their sum in the limit where $N_c$ goes to infinite represents the continuous Green's function. The plotted spectra are created by replacing the sum over delta functions by a sum over Lorenzians. The spectra in the top row are a sum of Lorenzians with a full width at half maximum of $\Gamma = 0.01$, the spectra in the bottom row are created from a sum of Lorenzians with a full width at half maximum of $\Gamma = 0.1$. From left to right we show calculations including 3, 11, 31, 101 and 301 bath sites. Each panel shows calculations for $U=0$ to 2 in steps of 0.25 in units of the band width.

For $U$ equal to zero, the impurity Green's function has exactly the same number of poles as the bath Green's function. For large $U$, the number of poles in either the upper or the lower Hubbard band is, again, roughly equal to the number of poles in the bath Green's function, although the total number of poles in the impurity Green's function, in principle, is allowed to be much larger. Numerically, it turns out that in the large $U$ limit, from the 1000 poles we include in the impurity Green's function, only a fraction, roughly equal to the number of poles in the bath Green's function, carries appreciable weight.

The calculations show a systematic convergence with increasing numbers of poles in the bath Green's function. For large and small values of $U$ the increase in number of poles enhances the spectral resolution. In order to get continuous spectra, one needs to broaden by a Lorentzian with full width at half maximum equal to three times the band width divided by the number of poles in the bath Green's function. The inclusion of 300 poles in the bath Green's function thus allows one to get a spectral resolution of 1\% of the band width.

Close to the metal-insulator transition there are substantial differences when the number of poles in the bath Green's function is enhanced. With only 3 poles in the bath Green's function, we find the metal-insulator transition to take place between $U=0.5$ and $U=0.75$. With 11 poles the transition takes place between $U=1.0$ and $U=1.25$. For 31 and 101 bath sites the transition takes place between $U=1.25$ and $U=1.5$. For 301 bath sites we even find a metallic solution for $U=1.50$.\cite{Karski:2005cm} In principle, there is a large range of values of $U$ where one can find both a metallic and an insulating solution. The calculations here always started from a metallic bath Green's function. When both solutions are possible we show the metallic solution. The fact that the metal-insulator transition is reduced in $U$ when fewer poles are included in the bath Green's function becomes clear if one looks at the approximations made. Due to the discretization of the bath Green's function, the system considered, in principle, always becomes an insulator, with a gap equal to the band width divided by the number of poles considered. The smaller the number of poles considered, the larger is the gap in the bath Green's function. Coulomb repulsion enhances this gap. The enhancement of the gap due to correlations is more effective if one already starts with a reasonably large gap for the uncorrelated system.

\subsection{Comparison to literature}

\begin{figure}
    \includegraphics[width=0.45\textwidth]{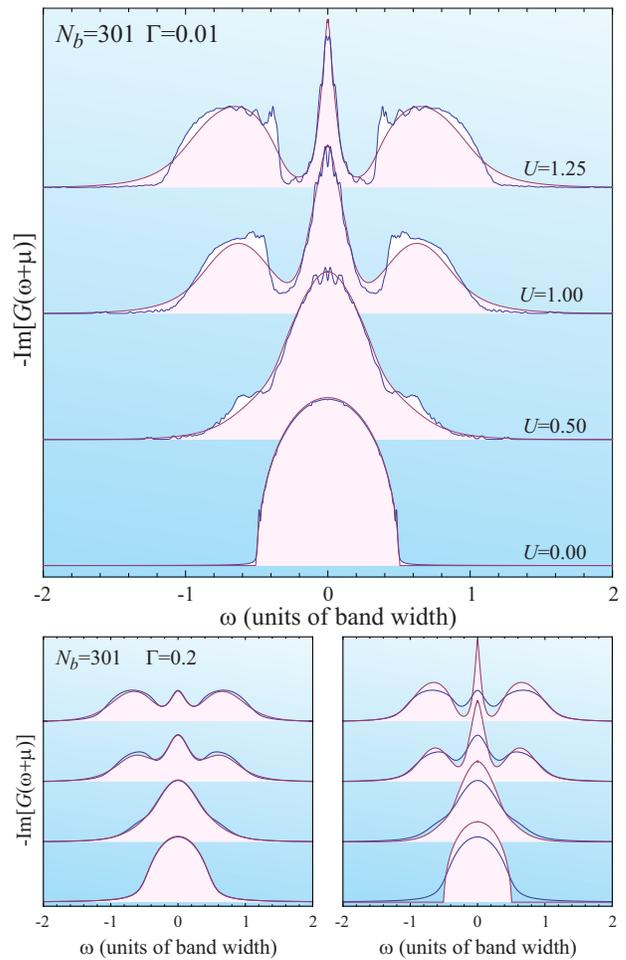}
    \caption{(Color online) Comparison between the NRG results as obtained by Bulla \textit{et al.}\cite{Bulla:1999js,  Bulla:2001uj, Bulla:2005en, Bulla:2006ek, Bulla:2008jn} (solid curves with red thin lines) and our calculations (thick blue lines) for $U=0.0$, 0.5, 1.0 and 1.25. Top panels show a Lorentzian broadening of full width at half maximum of 0.01, bottom panels show a Lorentzian broadening of full width at half maximum of 0.2 on both the ED and NRG results (left) or only the ED results (right).}
    \label{Fig:NRGBulla}
 \end{figure}

\begin{figure*}
    \includegraphics[width=1.00\textwidth]{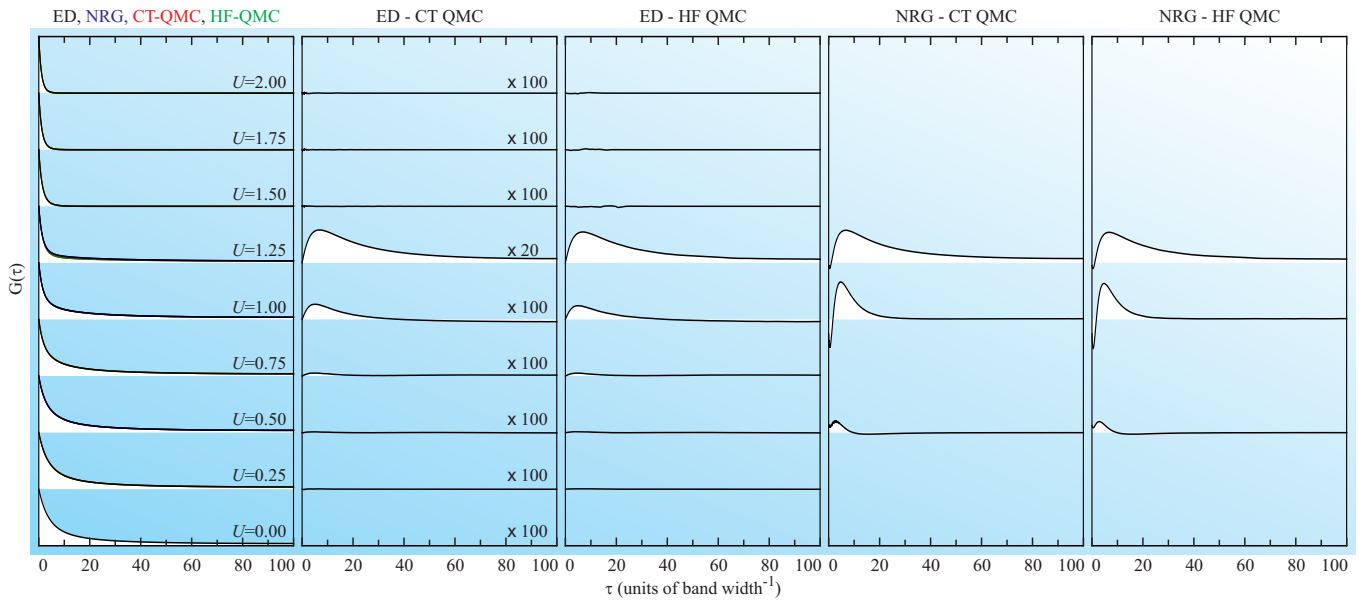}
    \caption{(color online) Comparison between the QMC calculations obtained with the CT (\textit{TRIQS} package\cite{Wernet:2006ko, Wernet:2006iz, Boehnke:2011dd}) or HF algorithm and the ED or NRG (Bulla \textit{et al.}\cite{Bulla:1999js, Bulla:2001uj, Bulla:2005en, Bulla:2006ek, Bulla:2008jn}) results for $U=0.0$, to $U=2.0$ in steps of $0.25$. From left to right we show $G(\tau)$ calculated with the four different methods, the difference between our method and CT-QMC, the difference between our method and HF-QMC, the difference between NRG and CT-QMC and the difference between NRG and HF-QMC. The difference plots are multiplied by a factor of 100 (20 for $U=1.25$) compared to the plots in the left panel.}
    \label{Fig:GTau}
 \end{figure*}

\begin{table*}
\begin{tabular}{c|ccc|r@{.}lr@{.}lr@{.}l|r@{.}lr@{.}lr@{.}l}
 $\Sigma_c(\omega)$ & \multicolumn{3}{c|}{Analytical} & \multicolumn{6}{c|}{$N_b=3$}& \multicolumn{6}{c}{$N_b=301$}\\
  $U$ & $U/2$ & $U^2/4$ & $U^3/8$ & \multicolumn{2}{c}{$M_{\Sigma}^{(-1)}$} & \multicolumn{2}{c}{$M_{\Sigma}^{(0)}$} & \multicolumn{2}{c|}{$M_{\Sigma}^{(1)}$} & \multicolumn{2}{c}{$M_{\Sigma}^{(-1)}$} & \multicolumn{2}{c}{$M_{\Sigma}^{(0)}$} & \multicolumn{2}{c}{$M_{\Sigma}^{(1)}$}  \\ \hline
 0.00 & 0.0000 & 0.0000 & 0.0000 & 0&0000 & 0&0000 & 0&0000 & 0&0000 & 0&0000 & 0&0000 \\
   &   &   &   & $-4$&$9\times 10^{-31} $&$ 7$&$1\times 10^{-16} $&$ -4$&$7\times 10^{-19} $&$ 7$&$9\times 10^{-19} $&$ 6$&$9\times 10^{-10} $&$ 5$&$8\times 10^{-10} $\\
 0.25 & 0.1250 & 0.0156 & 0.0020 & 0&1250 & 0&0156 & 0&0020 & 0&1250 & 0&0156 & 0&0019 \\
   &   &   &   & 0& & $1$&$9\times 10^{-16}$ & $1$&$8\times 10^{-17}$ & 0& & $-2$&$8\times 10^{-5}$ & $-3$&$5\times 10^{-6}$ \\
 0.50 & 0.2500 & 0.0625 & 0.0156 & 0&2500 & 0&0625 & 0&0156 & 0&2500 & 0&0625 & 0&0156 \\
   &   &   &   & $1$&$1\times 10^{-16}$ & $3$&$6\times 10^{-16}$ & $9$&$0\times 10^{-17}$ & 0& & $-3$&$6\times 10^{-5}$ & $-9$&$2\times 10^{-6}$ \\
 0.75 & 0.3750 & 0.1406 & 0.0527 & 0&3750 & 0&1406 & 0&0527 & 0&3750 & 0&1406 & 0&0527 \\
   &   &   &   & $1$&$1\times 10^{-16}$ & $2$&$7\times 10^{-17}$ & $1$&$3\times 10^{-16}$ &  0& & $-5$&$4\times 10^{-5}$ & $-2$&$0\times 10^{-5}$ \\
 1.00 & 0.5000 & 0.2500 & 0.1250 & 0&5000 & 0&2500 & 0&1250 & 0&5000 & 0&2500 & 0&1250 \\
   &   &   &   & 0& & $8$&$3\times 10^{-16}$ & $5$&$8\times 10^{-16}$ & $2$&$2\times 10^{-16}$ & $-2$&$1\times 10^{-5}$ & $-1$&$0\times 10^{-5}$ \\
 1.25 & 0.6250 & 0.3906 & 0.2441 & 0&6250 & 0&3906 & 0&2441 & 0&6250 & 0&3907 & 0&2442 \\
   &   &   &   & 0& & $-7$&$7\times 10^{-16}$ & $-5$&$0\times 10^{-16}$ & 0& & $4$&$5\times 10^{-5}$ & $2$&$8\times 10^{-5}$ \\
 1.50 & 0.7500 & 0.5625 & 0.4219 & 0&7500 & 0&5625 & 0&4219 & 0&7500 & 0&5625 & 0&4219 \\
   &   &   &   & 0& & $1$&$7\times 10^{-15}$ & $1$&$1\times 10^{-15}$ & 0& & $-9$&$9\times 10^{-6}$ & $-7$&$4\times 10^{-6}$ \\
 1.75 & 0.8750 & 0.7656 & 0.6699 & 0&8750 & 0&7656 & 0&6699 & 0&8750 & 0&7656 & 0&6699 \\
   &   &   &   & 0& & $-2$&$3\times 10^{-15}$ & $-8$&$8\times 10^{-16}$ & 0& & $-3$&$7\times 10^{-6}$ & $-3$&$2\times 10^{-6}$ \\
 2.00 & 1.0000 & 1.0000 & 1.0000 & 1&0000 & 1&0000 & 1&0000 & 1&0000 & 1&0000 & 1&0000 \\
   &   &   &   & 0& & $-1$&$5\times 10^{-15}$ & $-1$&$1\times 10^{-15}$ & 0& & $-5$&$5\times 10^{-6}$ & $-5$&$5\times 10^{-6}$ \\ \hline\hline
 $G_c(\omega)$ & \multicolumn{3}{c|}{Analytical} & \multicolumn{6}{c|}{$N_b=3$}& \multicolumn{6}{c}{$N_b=301$}\\
  $U$ & $U/2$ & $1/16+U^2/2$ & $3U/32+U^3/2$ & \multicolumn{2}{c}{$M_{G_c}^{(1)}$} & \multicolumn{2}{c}{$M_{G_c}^{(2)}$} & \multicolumn{2}{c|}{$M_{G_c}^{(3)}$} & \multicolumn{2}{c}{$M_{G_c}^{(1)}$} & \multicolumn{2}{c}{$M_{G_c}^{(2)}$} & \multicolumn{2}{c}{$M_{G_c}^{(3)}$}  \\ \hline
0.00 & 0.0000 & 0.0625 & 0.0000 & 0&0000 & 0&0625 & 0&0000 & 0&0000 & 0&0625 & 0&0000 \\
   &   &   &   & $-4$&$9\times 10^{-31}$ & $7$&$2\times 10^{-16}$ & $2$&$7\times 10^{-30}$ & $6$&$6\times 10^{-18}$ & $-1$&$6\times 10^{-7}$ & $2$&$9\times 10^{-9}$ \\
 0.25 & 0.1250 & 0.0938 & 0.0313 & 0&1250 & 0&0938 & 0&0313 & 0&1250 & 0&0937 & 0&0312 \\
   &   &   &   & 0& & $2$&$2\times 10^{-16}$ & $6$&$9\times 10^{-17}$ & 0& & $-2$&$5\times 10^{-5}$ & $-9$&$6\times 10^{-6}$ \\
 0.50 & 0.2500 & 0.1875 & 0.1094 & 0&2500 & 0&1875 & 0&1094 & 0&2500 & 0&1875 & 0&1093 \\
   &   &   &   & $1$&$1\times 10^{-16}$ & $4$&$7\times 10^{-16}$ & $3$&$5\times 10^{-16}$ & 0& & $-3$&$6\times 10^{-5}$ & $-2$&$7\times 10^{-5}$ \\
 0.75 & 0.3750 & 0.3438 & 0.2813 & 0&3750 & 0&3438 & 0&2813 & 0&3750 & 0&3437 & 0&2812 \\
   &   &   &   & $1$&$1\times 10^{-16}$ & $1$&$1\times 10^{-16}$ & $2$&$2\times 10^{-16}$ & $1$&$1\times 10^{-16}$ & $-4$&$8\times 10^{-5}$ & $-5$&$4\times 10^{-5}$ \\
 1.00 & 0.5000 & 0.5625 & 0.5938 & 0&5000 & 0&5625 & 0&5938 & 0&5000 & 0&5625 & 0&5937 \\
   &   &   &   & 0& & $8$&$9\times 10^{-16}$ & $1$&$6\times 10^{-15}$ & $1$&$1\times 10^{-16}$ & $-2$&$6\times 10^{-5}$ & $-3$&$8\times 10^{-5}$ \\
 1.25 & 0.6250 & 0.8438 & 1.0938 & 0&6250 & 0&8437 & 1&0937 & 0&6250 & 0&8438 & 1&0938 \\
   &   &   &   & 0& & $-6$&$7\times 10^{-16}$ & $-1$&$3\times 10^{-15}$ & $1$&$1\times 10^{-16}$ & $4$&$2\times 10^{-5}$ & $7$&$8\times 10^{-5}$ \\
 1.50 & 0.7500 & 1.1875 & 1.8281 & 0&7500 & 1&1875 & 1&8281 & 0&7500 & 1&1875 & 1&8281 \\
   &   &   &   & 0& & $1$&$8\times 10^{-15}$ & $4$&$0\times 10^{-15}$ & $1$&$1\times 10^{-16}$ & $-1$&$1\times 10^{-5}$ & $-2$&$4\times 10^{-5}$ \\
 1.75 & 0.8750 & 1.5938 & 2.8438 & 0&8750 & 1&5937 & 2&8437 & 0&8750 & 1&5937 & 2&8437 \\
   &   &   &   & 0& & $-2$&$2\times 10^{-15}$ & $-4$&$9\times 10^{-15}$ & 0& & $-3$&$5\times 10^{-6}$ & $-9$&$2\times 10^{-6}$ \\
 2.00 & 1.0000 & 2.0625 & 4.1875 & 1&0000 & 2&0625 & 4&1875 & 1&0000 & 2&0625 & 4&1875 \\
   &   &   &   & 0& & $-1$&$8\times 10^{-15}$ & $-4$&$4\times 10^{-15}$ & 0& & $-6$&$3\times 10^{-6}$ & $-1$&$9\times 10^{-5}$ \\
\end{tabular}
\caption{Comparison of the analytical and numerical moments of the impurity self-energy $\Sigma_c(\omega)$ (top panel) and the impurity Green's function $G_c(\omega)$ (bottom panel) for 3 ($N_b=3$) and 301 ($N_b=301$) bath sites and different values of $U$. The zeroth and first moment of the $G_c(\omega)$ are exactly equal to 0 and 1 for all $U$ and number of bath orbitals. The even rows show the value of the moment, the odd rows show the difference between the numerical and analytical value.}
    \label{Table:Moments}
\end{table*}

The calculations of the dynamical mean-field solution of the Hubbard model on the Bethe lattice can be compared to a huge amount of literature data. We here include three examples explicitly. For the metallic cases we compare the ED to the NRG results as obtained by Bulla \textit{et al.}\cite{Bulla:1999js,  Bulla:2001uj, Bulla:2005en, Bulla:2006ek, Bulla:2008jn}. NRG in this case is a highly efficient method and the comparison thus provides a stringent test on the current method. We furthermore compare our ED to QMC calculations. We used both the HF algorithm as well as the CT algorithm as implemented in the \textit{TRIQS} package.\cite{Wernet:2006ko, Wernet:2006iz, Boehnke:2011dd} In order to avoid the analytical continuation of the QMC spectra from the imaginary to the real axis, we transformed our results to the imaginary time axis. In the third subsection we compare to analytically known sum rules for the Green's function and self-energy of a Hubbard model on a Bethe lattices solved within the DMFT approximation. 

\subsubsection{Comparison to NRG results}

In figure \ref{Fig:NRGBulla} we show a comparison between our results obtained with ED and the results obtained by Bulla \textit{et al.}\cite{Bulla:1999js,  Bulla:2001uj, Bulla:2005en, Bulla:2006ek, Bulla:2008jn} using NRG. We find the position and weight of the upper Hubbard band, the lower Hubbard band, and the quasiparticle peak to be extremely similar. However, there are evidently two differences. 

First, the ED results show extra wiggles, almost like noise, compared to the NRG calculations. Such extra features have been reported before, but no full interpretation nor understanding exists. \cite{Karski:2005cm, Karski:2008hc, Garcia:2007de, Miranda:2008cs, Krivenko:2012hn, Zitko:2009jq} It has been shown that for an antiferromagnetic solution the upper and lower Hubbard bands show magnon sidebands.\cite{Sangiovanni:2006ez} For the paramagnetic solution it is not obvious that these features (paramagnon sidebands) should exist as well. In our calculations these wiggles are most probably related to numerical instabilities in the Lanczos algorithm. The use of iterative schemes including Lanczos, as well as the use of tridiagonal matrices to represent the Green's function, can lead to numerical instabilities when using finite precision math. This is not just a problem of ED, but is a numerical challenge for any method using a Krylov basis set on which the Hamiltonian is tridiagonal. All of these methods should take care to prevent number loss within the algorithm when creating the Krylov basis.

Second, the ED results are sharper at the high-energy side of the upper and lower Hubbard bands. These spectra still have a tail that decays for $\omega \to \infty$, but with much smaller spectral weight. The NRG results are obtained on a logarithmic mesh; therefore, the accuracy close to the Fermi energy is higher than the accuracy of the Hubbard bands. In practice, this can be overcome by an additional broadening at higher frequencies. If one compares the NRG results to our results broadened by a Lorentzian of full width at half maximum of 0.2 the agreement at the Hubbard bands is perfect, as can be seen in the bottom panels of figure \ref{Fig:NRGBulla}. The overall agreement between our ED results and the NRG results is considerably good.

\subsubsection{Comparison to QMC results}

In order to further compare our numerical results, we performed QMC calculations. They are performed at an inverse temperature of $\beta = 200$ in units of the band-width of $G_0$. The spin-up and spin-down Green's functions are averaged in order to force a paramagnetic solution. The HF\cite{Gunnarsson:2010jt, Merino:2012fc} calculations use 1600 steps in $\beta$ for $1.0\leq U \leq 2.0$ and 1200 steps for $0.0\leq U \leq 0.75$. In the case of CT QMC calculations, 10000 $\tau$~points (1025 Matsubara frequencies) were used to sample $G(\tau)$ [$G(\mathrm{i}\omega)$], respectively. For both the HF and CT QMC, it was ensured that the Green's function obey the correct asymptotic behavior (noise reduction of the numerical data). The ED and NRG results are obtained at $\beta \rightarrow \infty$, i.e. at 0 Kelvin, but the QMC ones are obtained at finite temperature. The former Green's function is the ground state expectation value, whereas the latter represents the statistical average at finite temperature, which does lead to differences in the metallic regime close to the metal insulator transition. In order to transform the real-frequency results to the imaginary time axis, we included a fictitious temperature ($\beta_f = 200$) in the transformation.

In the left panel of figure~\ref{Fig:GTau} we show our ED results, the QMC results and the NRG results for $U=0$ to $U=2$ in steps of 0.25. They seem to agree well. (Note that one cannot distinguish the four lines plotted in the left panel of figure~\ref{Fig:GTau}) In imaginary time Green's functions the spectral complexity is encoded in the fine details. Hence, one should compare the differences between the three Green's functions obtained by different methods in more detail. In the right four panels of figure \ref{Fig:GTau} we show the difference between $G(\tau)$ calculated with (1) ED and CT-QMC, (2) ED and HF-QMC, (3) NRG and CT-QMC and (4) NRG and HF-QMC. One should first observe that up to the statistical accuracy with which the QMC calculations have been preformed the HF and the CT algorithm give the same results. For the metallic cases the differences between our ED calculations and QMC calculations become larger if one gets closer to the metal-to-insulator transition. The same behavior is true for the comparison between NRG and QMC. This is not related to numerical problems in either of the two methods, but to the fact that the QMC calculations are preformed at finite temperature ($\beta = 200$), whereas the ED and the NRG results have been obtained at exactly 0 K. The critical $U_c$ decreases with temperature up to the critical point; hence, at finite temperature the metal-to-insulator transition occurs for lower $U$ values than at $T=0$ K.\cite{Terletska:2011fp} In fact, with increasing $U$ one notices, that in QMC the spectral weight at the Fermi level in the metallic regime gets smaller than in ED/NRG [$G(\beta/2) \propto A(w=0)$]. For the insulating case we basically find, up to the statistical accuracy with which the QMC calculations are preformed, agreement between all different methods shown. Comparing with QMC and NRG we find that the method works well. Note that the small differences between QMC and NRG at $\tau=0$ are due to the coarse mesh of the NRG data at large $\omega$, which introduce problems in the transformation from real frequency to imaginary time.

\subsubsection{Comparison to analytical moment and bath hybridization sumrules}

Several analytical sum rules exist that relate the first four moments of the impurity Green's function, the first two moments, and an additional constant of the self-energy, as well as the first four moments of the bath Green's function, to analytically known expressions.\cite{Potthoff:1997wv, Herrmann:1997wi, Potthoff:1998vr, Potthoff:1998to, Wegner:1998ux, Blumer:2003tz} Furthermore, Koch \textit{et al.}\cite{Koch:2008bx} showed that the total hybridization between the impurity and the bath is related to the first and second moments of the noninteracting Green's function. Their relations in our present notation become particularly transparent. Given the noninteracting Green's function $G_0(\omega)$ as defined in equation (\ref{eq:G0}) and the bath Green's function $G_b(\omega)$ as defined in equation (\ref{eq:Gb}) the hybridization sum rule states that $a_1$ of the noninteracting Green's function is equal to $a_1^b$ of the bath Green's function. Our implementation of the self-consistency loop as shown in section \ref{section:loop} and particularly equation (\ref{eq:bathnew}) guarantees that this sum rule is exactly fulfilled. The momentum sum rules need to be checked agains their numerical values. They are valid for continuous Green's functions and self-energy, but it is \textit{a-priory} not obvious how the discretization used in the method presented here influences the moments of the Green's functions. The reduction of poles as described in Appendix \ref{app:polered} as well as numerical instabilities, could, in principle lead to a violation of these sum rules. Below we show that the moment sum rules are fulfilled very well with the method presented in this paper.

The moments of a Green's function (or self-energy) are defined as
\begin{equation}
M^{(m)}_G = \frac{1}{\pi} \int_{-\infty}^{\infty} -\mathrm{Im}[G(\omega)] \omega^m d \omega.
\end{equation}
Direct numerical evaluation of this integral is difficult due to problems with number loss. One can rewrite this integral to a series expansion in $1/\omega$ whose expansion coefficients are given by $M^{(m)}_G$.\cite{Potthoff:1998vr} With the use of the Kramers-Kronig relations,
\begin{equation}
G(\omega) = \frac{\mathrm{i}}{\pi} \int_{-\infty}^{\infty} \frac{G(\omega')}{\omega-\omega'} d \omega',
\end{equation}
one can rewrite the Green's function as a series expansion in $1/\omega$,
\begin{equation}
G(\omega) = \sum_{m=0}^{\infty} \frac{M^{(m)}_G}{\omega^{m+1}}.
\end{equation}

The Green's functions in our method are represented by a sum over poles as
\begin{equation}
G(\omega) = \sum_{i} \frac{\beta_i^2}{\omega-\alpha_i}.
\end{equation}
In order to calculate the moments of this Green's function we create a Laurent series of $G(\omega)$:
\begin{equation}
G(\omega) = \sum_{m=0}^{\infty} \sum_{i} \frac{\beta_i^2 \alpha_i^{m}}{\omega^{m+1}}.
\end{equation}
The moments of the Green's function can thus be expressed in term of $\alpha_i$ and $\beta_i$ which are used as numerical values to store the Green's function:
\begin{equation}
M^{(m)}_G = \sum_{i} \beta_i^2 \alpha_i^{m}.
\end{equation}

The analytical expressions for the moments of the Green's function of a one band Hubbard model on a Bethe lattice with $W=1$, solved within the DMFT approximation, are
\begin{align}
G_c(\omega) &= \frac{1}{\omega^1} + \frac{U/2}{\omega^2} + \frac{1/16+U^2/2}{\omega^3} \\
\nonumber   &+ \frac{3U/32+U^3/2}{\omega^4} + \mathcal O\left(\frac{1}{\omega}\right)^5,\\
\nonumber   M^{(0)}_{G_c} &=1,\\
\nonumber   M^{(1)}_{G_c} &=U/2,\\
\nonumber   M^{(2)}_{G_c} &=1/16+U^2/2,\\
\nonumber   M^{(3)}_{G_c} &=3U/32+U^3/2.
\end{align}
For the self-energy they are
\begin{align}
\Sigma_c(\omega) &= \frac{U/2}{\omega^0} + \frac{U^2/4}{\omega^1} + \frac{U^3/8}{\omega^2} + \mathcal O\left(\frac{1}{\omega}\right)^3,\\
\nonumber   M^{(-1)}_{\Sigma_c} &=U/2,\\
\nonumber   M^{(0)}_{\Sigma_c} &=U^2/4,\\
\nonumber   M^{(1)}_{\Sigma_c} &=U^3/8,
\end{align}
where we defined $M^{(-1)}_{\Sigma_c}$ as the prefactor of $1/\omega^0$ in the series expansion in $1/\omega$.

In Table \ref{Table:Moments} we compare our numerical results with the analytical values. The top half of the table shows the moments of the self-energy; the bottom half shows the moments of the impurity Green's function. The calculations are done for $U=0$ to $U=2.0$ in steps of $0.25$ the same steps as used for the spectra shown in figure \ref{Fig:Results}. We show calculations for $N_b=3$ and $N_b=301$. The even rows show the analytical and numerical moments, the odd rows show the difference between the numerical and analytical values. The first moment of the Green's function is not included as this is exactly equal to 1 for all calculations. We find that already for three bath sites there is perfect (down to the numerical precision possible in a computer $\sim 10^{-16}$) agreement between our numerical and the analytical results. It might be surprising that the moments are represented so well, whereas the spectra (see the left panels of figure \ref{Fig:Results}) are not converged in the number of bath sites: For $N_b=3$ we find roughly two peaks per Hubbard band and a transition to the insulating state at much too low values of $U$. This shows once again that the moments of a Green's function can be used as a criteria to falsify a numerical method, but even if a numerical method has several moments of the Green's function correct it does not imply that the method works. It might come as a surprise that the moments of the Green's function are better reproduced with $N_b=3$ (15 digits correct) than with $N_b=301$ (5 digits correct). This is most probably related to number loss in the calculations, which is a larger concern when more bath states are included and also probably the reason why increasing the basis to include 1001 bath orbitals did not improve the spectral function further. In any case, we can conclude that the analytically known values for the first four moments of the Green's function and the two moments of the self-energy as well as the constant offset in the self-energy are well reproduced in our method.

\section{Conclusion}

In this paper we present an efficient ED-based real-frequency solver for the general Anderson impurity problem and DMFT. It alleviates the exponential increasing Hilbert space encountered by conventional ED algorithms as a function of the number of bath sites. A specific bath geometry is realized upon which basis set optimization can be applied. The restricted Hilbert space allows calculations including a few hundred bath sites at moderate cost, which solve for spectral functions with energy resolution better than $1/\mathcal O(10^2)$ of the bandwidth. Good agreement with other methods including NRG, HF-QMC, and CT-QMC is obtained for model systems over a wide parameter space.

We would like to thank Silke Biermann, Philipp Hansmann, Alessandro Toschi, Giorgio Sangiovanni and Karsten Held for stimulating discussions. Financial support from the Deutsche Forschungsgemeinschaft through Grant No. FOR 1346 is gratefully acknowledged.

\appendix

\section{Notation}
\label{app:notation}

In the main paper as well as in the appendixes we use $\tau$ as an index for the different fermion quantum states (spin, orbital, site) within the impurity. $N_{\tau}$ is the total number of these degrees of freedom. In most equations the sum over $\tau$ is suppressed. For example, let $\alpha$ be an $N_{\tau}$ by $N_{\tau}$ matrix with elements $\alpha_{\tau,\tau'}$, then
\begin{equation}
\alpha a^{\dag}_{\phantom{\tau}} a^{\phantom{\dag}}_{\phantom{\tau}} \equiv \sum_{\tau,\tau'}^{N_{\tau},N_{\tau}} \alpha_{\tau,\tau'} a^{\dag}_{\tau} a^{\phantom{\dag}}_{\tau'}
\end{equation}
The same notation and suppression of internal degrees of freedom are used for the bath sites.

The sum of a scalar and a matrix is used as a shorthand for the sum of a scalar times the identity matrix. The inverse of a matrix is given by a fraction and the resolvent of a matrix is assumed to be taken such that the poles are in quadrants III and IV. In formula this is
\begin{equation}
\frac{1}{\omega-\alpha} \equiv \lim_{\eta\to 0^+} \frac{1}{\omega \mathbb{I} - \alpha + \mathrm{i} \eta},
\end{equation} 
with $\mathbb{I}$ an $N_{\tau}$-by-$N_{\tau}$ identity matrix and $\alpha$ a general $N_{\tau}$-by-$N_{\tau}$ matrix.

The square of a matrix, divided by another matrix should be read as the product of three matrices:
\begin{equation}
\frac{\beta^2}{\omega-\alpha} \equiv \beta^{\dag} \frac{1}{\omega-\alpha} \beta^{\phantom{\dag}},
\end{equation}
with $\alpha$ and $\beta$ $N_{\tau}$-by-$N_{\tau}$ matrices. 

We define \textit{sites} as a set of one-electron states that arise from the quantization of the bath Green's function. The term \textit{site} is chosen because for a finite size tight binding lattice model this quantization can be taken to overlap with the real sites in the lattice model. As stated above, each site (including the impurity) can have additional degrees of freedom labeled by $\tau$. The impurity site is labeled by $i$, the bath sites are labeled by $b_j$ or by $b$, $v_j$ and $c_j$. one-electron states are defined by creating an electron at a given site: $a^{\dag}_i$ or $a^{\dag}_{b_j}$ for impurity or bath sites. We recombine bath sites to optimize our basis and bath geometry. The relation between the new ($b_{\varsigma}$) and old ($b_j$) sites is given by a unitary rotation matrix $U$ with element $u_{j;\varsigma}$ such that: $a^{\dag}_{b_{\varsigma}}=\sum_j u_{j;\varsigma} a^{\dag}_{b_{j}}$. A given filling of these sites defines a single Slater determinant function labeled by $\phi$. For $N$ electrons, the set of Slater determinants is given by all subsets $D_i$ of length $N$ of the possible fermions ($\tau$) at either the impurity ($i$) or the bath ($b_j$) sites:
\begin{equation}
|\phi_i\rangle=\Pi_{\gamma\in D_i} a^{\dag}_{\gamma} |0\rangle.
\end{equation}
The operator $a^{\dag}_{\gamma}$ creates a single electron with quantum numbers ($\tau$, $i$, $b_j$) indexed by $\gamma$. The Slater determinant $\phi_i$ thus represents a state with $N$ electrons. Given a set of Slater determinants, one can define the ground state $\psi$ as a linear combination of these many electron determinants:
\begin{equation}
\psi = \sum_i \alpha_i \phi_i,
\end{equation}
with $\alpha_i$ numerical factors defining the state and $\sum_i |\alpha_i|^2=1$ to normalize the state.
We use, generally, $\psi$ to label a many-Slater-determinant eigenstate on a given basis and $\varphi$ to label a many-Slater-determinant basis state, which is part of the Krylov basis of the Hamiltonian starting from a specific state.

\section{Transformations between different representations of the Green's function}
\label{app:Gtrans}

In this paper the Green's function (and self-energy) is expressed as an analytical function involving the sum over $\alpha_i$ and $\beta_i$, with $\beta_i$ related to the spectral weight and $\alpha_i$ related to the energy of the poles. We use different representations of the Green's function in different parts of the code. The Lanczos algorithm produces the Green's function as a continued fraction, equation (\ref{eq:contfrac}). The DMFT self-consistency loop is written using the Green's function as a sum over poles, equation (\ref{eq:sumoverpoles}), and as the inverse of the sum over several poles, equation (\ref{eq:stargeometry}). In all cases the Green's function can be represented as the resolvent of a matrix ($H$) and transformations between the different representations of the Green's function are unitary matrix transformations of this matrix. The basis of the matrix $H$ can be interpreted as sites and $H$ as the Hamiltonian determining the onsite energy and hopping of a single electron between different sites. 

The Green's functions are always represented as a set of delta functions. Only when the Green's function is plotted, after the full self-consistency is reached, we broaden the Green's function. (Replace the sum over delta functions by a sum over Lorentzians.) In figure \ref{Fig:Results} we show two different Lorentzian broadenings [full with half maximum of 0.01 (top) and 0.1 (bottom)]. The transformation to the imaginary axis as shown in figure \ref{Fig:GTau} is done without a broadening on the Green's function. The transformation from the real to imaginary axis involves an integral of a kernel times the Green's function, which can be performed straightforwardly when the Green's function is given as a list of delta functions at energy $\alpha_i$ and weight $\beta_i^2$.

In this section we discuss the transformations between the different representations in more detail. If one starts from a density functional theory calculation, the noninteracting Green's function is often only known by the spectral function or density of states represented by a list of energies and intensities ($A_{k,\omega_{i,k}}$). This defines the Green's function as:
\begin{equation}
G(\omega) = \lim_{\eta \to 0^+}\sum_{k,i} \frac{A_{k,\omega_{i,k}}}{\omega-\omega_{i,k}+\mathrm{i}\eta}.
\end{equation}
Combining the sum over momenta ($k$) and quantized energies $\omega_{i,k}$ into a single sum and rewriting the numerical parameters as $\alpha_i$ and $\beta_i$ we get:
\begin{equation}
\label{eq:sumoverpoles}
G(\omega) = \sum_{i=1}^{N} \frac{\beta_i^2}{\omega-\alpha_i}.
\end{equation}
We would like to find a matrix whose resolvent is equal to this Green's function, such that numerical operators on the Green's function can be implemented as matrix operations. In order to do this we define the matrix
\begin{equation}
\label{eq:HDiagonal}
H_{e}=\left(
\begin{array}{ccccc}
\alpha_1 & 0       & 0      & 0      & 0        \\
0        &\alpha_2 &0       & 0      & 0        \\
0        & 0       & \ddots & 0      & 0        \\
0        & 0       & 0      & \ddots & 0        \\
0        & 0       & 0      &0       &\alpha_N 
\end{array}
\right),
\end{equation}
and the vector
\begin{equation}
\label{eq:chi0}
\chi_0=\{\beta_1,\beta_2,\hdots,\beta_{N-1},\beta_N\}.
\end{equation}
Using $H_e$ and $\chi_0$, the Green's function is defined as the inner product of $\chi_0$ and the resolvent of $H_e$:
\begin{equation}
G(\omega)=\left\langle \chi_0 \left| \frac{1}{\omega-H_e} \right| \chi_0 \right\rangle.
\end{equation}
Transformations between different representations of the Green's function as shown for example in figure \ref{Fig:Structure} can now be written as matrix transformations on $H_e$:
\begin{equation}
G(\omega)=\left\langle U \chi_0 \left| \frac{1}{\omega-U^{\dag}H_eU} \right| U \chi_0 \right\rangle.
\end{equation}
In order to define the different unitary transformations $U$ that relate the Green's function in the representation as shown in equation (\ref{eq:sumoverpoles}) to the Green's function in the tridiagonal or Anderson representation as depicted in figure \ref{Fig:Structure} we take two steps. In the first step we create $H_1$, which is a dense matrix whose top left-most element of the resolvent represents the Green's function. In the second step we apply a unitary matrix transformation on the elements $2$ to $N$ of $H_1$ to obtain the Green's function represented by a tridiagonal ($H_t$) or Anderson ($H_A$) Hamiltonian. In the first step we define $U_1$ such that
\begin{equation}
\nonumber H_1 = U_1^{\dag} H_e U_1,
\end{equation}
with
\begin{equation}
\nonumber U_1 \chi_0 = \{1,0,0,\hdots,0\},
\end{equation}
and
\begin{align}
\nonumber G(\omega)=&\left\langle \{1,0,\hdots,0\} \left| \frac{1}{\omega-H_1} \right| \{1,0,\hdots,0\} \right\rangle\\
=&\left( \omega-H_1 \right)^{-1}_{[1,1]},
\end{align}
whereby the exponent in the last equation represents a matrix inversion and the subscript $[1,1]$ represents the element at position 1 after the matrix inversion. $H_1$ is a dense matrix of dimension $N$ by $N$ with $N$ equal to the number of poles in the Green's function as defined in equation (\ref{eq:sumoverpoles}). The elements of $H_1$ are given as
\begin{equation}
\label{eq:H1}
h^{(1)}_{i,j} = \langle \chi_j | H_e | \chi_i \rangle,
\end{equation}
with $\chi_0=\{\beta_1,\beta_2,\hdots,\beta_{N-1},\beta_N\}$ as defined in equation (\ref{eq:chi0}) and $\chi_i$ for $1 \leq i < N$ obtained from a Gram-Schmidt orthonormalization of a set of unit vectors that span the basis of $H_e$.

The Anderson representation of the Green's function where the site under consideration interacts with $N-1$ noninteracting other sites is given by the Hamiltonian
\begin{equation}
\label{eq:HAnd}
H_A=\left(
\begin{array}{ccccc}
\alpha^A_1    &\beta^A_1  &\beta^A_2& \hdots &\beta^A_{N-1}\\
\beta^A_1     &\alpha^A_2 & 0      & 0      & 0        \\
\beta^A_2     & 0        & \ddots & 0      & 0        \\
\vdots       & 0        & 0      & \ddots & 0        \\
\beta^A_{N-1}   & 0        & 0      & 0      &\alpha^A_{N} 
\end{array}
\right),
\end{equation}
and the corresponding Green's function is given as:
\begin{equation}
\label{eq:stargeometry}
G(\omega)=\frac{1}{\omega - \alpha^A_1 -\sum_{i=1}^{N-1} \frac{{\beta^A_i}^2}{\omega-\alpha^A_{i+1}} }.
\end{equation}

The unitary transformation relating the Hamiltonian $H_A$ in equation (\ref{eq:HAnd}) to $H_1$ as defined in equation (\ref{eq:H1}) is given by the eigenvectors of $H'_1$ with the elements of $H'_1$ defined as
\begin{equation}
h'^{(1)}_{i,j} = (1-\delta_{i,0})(1-\delta_{0,j}) h^{(1)}_{i,j}.
\end{equation}

A different unitary transformation of $H_1$ can lead to the representation of the Green's function where the site under consideration interacts with exactly one other site, which in turn interacts with one more site building a one dimensional chain of interactions. The Hamiltonian in this case is given by
\begin{equation}
H_t=\left(
\begin{array}{ccccc}
\alpha^t_1 &\beta^t_1  &0        & 0      & 0        \\
\beta^t_1  &\alpha^t_2 &\beta^t_2 & 0      & 0        \\
0         & \beta^t_2 & \ddots & \ddots & 0        \\
0         & 0        & \ddots & \ddots &\beta^t_{N-1}     \\
0         & 0        & 0      &\beta^t_{N-1} &\alpha^t_{N} 
\end{array}
\right),
\end{equation}
and the Green's function as
\begin{equation}
\label{eq:contfrac}
G(\omega)=\frac{1}{\omega - \alpha^t_1 - \frac{{\beta^t_1}^2}{ \omega - \alpha^t_2 - \frac{{\beta^t_2}^2}{\omega-\hdots  }}}
\end{equation}
$H_t$ is the tridiagonal form of $H_1$ whose elements are defined in equation (\ref{eq:H1}). This tridiagonal matrix is obtained by a standard Lanczos tridiagonalization routine. (See Appendix \ref{app:lanc} and references therein.)

\section{Lanczos}
\label{app:lanc}

There are several good review articles around describing the Lanczos algorithm.\cite{Weisse:2006go, Jaklic:1994ii} In general, we would not advise to implement the complete Lanczos routines itself, but to use one of the libraries available.\cite{Maschhoff:1996ve, Bergamaschi:2002vf} In this appendix we provide a short overview of the basic idea behind the Lanczos routine, which will help the reader in understanding the implementation of the Lanczos routines on a sparse, continuously optimized basis set. The Lanczos routines can be used to find the ground state of a large sparse matrix. Once the ground state is found the same routine can be used to calculate spectral functions, including the one particle Green's function. Here we provide some information on both procedures.

\subsection{Finding the ground state of a large sparse matrix}

The Hamiltonian $H$ can be represented on a basis as a large, sparse matrix. We can shift the onsite energy of this matrix such that all eigenvalues are negative. Next we define an arbitrary, random wave function $\varphi_0$. This wavefunction can be written as a linear combination of eigenstates,
\begin{equation}
\varphi_0 = \sum_i \alpha_i \psi_i,
\end{equation}
with $\psi_i$ eigenstates of H such that
\begin{equation}
H \psi_i = E_i \psi_i.
\end{equation}
The states $\psi_i$ are taken to be ordered such that
\begin{equation}
E_i \leq E_{i+1} < 0.
\end{equation}
The state $\psi_0$ is the ground state one would like to determine. The state $\varphi_1$ is defined by the recurrent relation
\begin{equation}
\label{eq:lancStep}
\varphi_{i+1} = \frac{H \varphi_i}{\sqrt{ \langle \varphi_i | H^2 | \varphi_i \rangle}}.
\end{equation}
Besides normalization, $\varphi_1$ is given by
\begin{equation}
\varphi_1 = \sum_i E_i \alpha_i \psi_i.
\end{equation}
As $|E_0| \geq |E_i|$ and $E_i<0$ $\forall i$ the overlap of $\varphi_1$ with the ground state $\psi_0$ is larger than the overlap of $\varphi_0$:
\begin{equation}
|\langle \varphi_1 | \psi_0 \rangle| \geq |\langle \varphi_0 | \psi_0 \rangle|.
\end{equation}
Repeatedly applying equation (\ref{eq:lancStep}) will lead to convergence of $\psi_i$ to the ground state: $\lim_{i\to\infty} \varphi_i = \psi_0$.

Although the above-described algorithm works and is extreme robust, convergence can be exponentially slow. In order to improve convergence, we define a Krylov space with a fraction of the size of the total Hamiltonian and diagonalize the matrix on this new basis.
Starting from a random vector $\varphi_0$, we define the Krylov basis by the recurrent relations:
\begin{align}
\label{eq:lancKryl}
\nonumber
\tilde{\tilde{\varphi}}_{i+1}&= H \varphi_i.\\
\nonumber
\tilde{\varphi}_{i+1}&=\tilde{\tilde{\varphi}}_{i+1} - \langle \varphi_i | \tilde{\tilde{\varphi}}_{i+1} \rangle \varphi_i - \langle \varphi_{i-1} | \tilde{\tilde{\varphi}}_{i+1} \rangle \varphi_{i-1}.\\
\varphi_{i+1} &= \frac{\tilde{\varphi}_{i+1}}{\sqrt{\langle \tilde{\varphi}_{i+1} | \tilde{\varphi}_{i+1} \rangle}}.
\end{align}
The first step defines the basis according to the idea that $H \varphi_i$ is closer to the ground state than $\varphi_i$. The second step assures that $\varphi_i$ is orthogonal to $\varphi_j$ for all $i\neq j$. The last step in equation (\ref{eq:lancKryl}) assures normalization of $\varphi_i$.

For large enough Krylov basis sets one can diagonalize the Hamiltonian in the Krylov basis, which is tridiagonal, and obtain the ground state of the full Hamiltonian. In practice, it works better to take moderately large Krylov basis sets (somewhere between 10 and 100) and obtain the ground state from the Hamiltonian in this basis. This function is then taken as the starting point for a new Krylov basis.\cite{Calvetti:1994wu, Wu:1999vb, Sundar:2000vy, Wu:2000vq, Kokiopoulou:2004bn} These steps are repeated until the state is converged to the ground state of the full Hamiltonian. A good mehtod to check the convergence is to test if
\begin{equation}
\label{eq:lancConv}
|\langle \psi_0 | H | \psi_0 \rangle|^2 = \langle \psi_0 | H^2 | \psi_0 \rangle.
\end{equation}

It is useful to note that numerical stability is an issue in this algorithm and numerical errors can build up, which should be dealt with using for example Kahan summation, additional orthogonalization, and restarting often enough.\cite{Calvetti:1994wu, Wu:1999vb, Sundar:2000vy, Wu:2000vq, Kokiopoulou:2004bn} In order to improve convergence and numerical stability one can shift the Hamiltonian such that not all eigenstates are negative, but the zero of energy is closer to the actual ground state energy. Furthermore, for systems with a large number of degenerate eigenstates it can be useful to use a block Lanczos algorithm where not one, but several eigenstates are created simultaneously. There is not one single strategy that works best for all Hamiltonians; therefore, implementations should change strategy when convergence becomes slow.

\subsection{Calculating spectral functions using Lanczos}

In order to calculate spectral or Green's functions, one needs to obtain the resolvent of the Hamiltonian projected to a particular state. In general,
\begin{equation}
g(\omega)=\lim_{\Gamma\to0^+} \left\langle \psi_0 \left| T^{\dag}_i \frac{1}{\omega-H+\mathrm{i}\frac{\Gamma}{2}} T^{\phantom{\dag}}_i \right| \psi_0 \right\rangle,
\end{equation}
with $T=a^{\phantom{\dag}}$ ($a^{\dag}$, $a^{\dag}_{\uparrow}a^{\phantom{\dag}}_{\downarrow}$, ...) for the Green's function related to photoemission (inverse photoemission, spin susceptibility, ...). 

We define
\begin{equation}
\varphi_0 = \frac{T^{\phantom{\dag}} \psi_0}{\sqrt{\langle \psi_0 | T^{\dag} T^{\phantom{\dag}} | \psi_0 \rangle}}
\end{equation}
and the Krylov basis by $\varphi_j$ as defined by the recurrence relations as given in equation (\ref{eq:lancKryl}). On this basis, the Hamiltonian ($H_{Krylov}$) is tridiagonal and can be parametrized by $\alpha_i$ and $\beta_i$:
\begin{equation}
H_{Krylov}=\left(
\begin{array}{ccccc}
\alpha_1    &\beta_1    & 0      & 0      & 0        \\
\beta_1   &\alpha_2    &\beta_2    & 0      & 0        \\
0      &\beta_2    & \ddots & \ddots & 0        \\
0      & 0      & \ddots & \ddots &\beta_{n}    \\
0      & 0      & 0      &\beta_{n}  &\alpha_{n+1} 
\end{array}
\right).
\end{equation}
The resolvent of a tridiagonal matrix is given as a continued fraction,
\begin{align}
\label{eq:contfracspec}
\nonumber
\left(
\begin{array}{ccccc}
\omega-\alpha_1    &-\beta_1    & 0      & 0      & 0        \\
-\beta_1   &\omega-\alpha_2    &-\beta_2    & 0      & 0        \\
0      &-\beta_2    & \ddots & \ddots & 0        \\
0      & 0      & \ddots & \ddots &-\beta_{n}    \\
0      & 0      & 0      &-\beta_{n}  &\omega-\alpha_{n+1} 
\end{array}
\right)^{-1}_{[1,1]}\\
=\frac{1}{\omega-\alpha_1-\frac{\beta_1^2}{\omega-\alpha_2-\frac{\beta_2^2}{\omega-\hdots}}},
\end{align}
which allows for a straight forward calculation of the Green's function corresponding to the transition operator $T$. 

For the calculation of spectral functions (as with the calculation of the ground state) one should be aware that the construct of the Krylov basis includes a fundamental numerical unstable algorithm. Additional orthonormalization steps can be mandatory in order to obtain correct results.

\section{Lanczos on a sparse basis}
\label{app:lancsparse}

 \begin{figure}
    \includegraphics[width=0.45\textwidth]{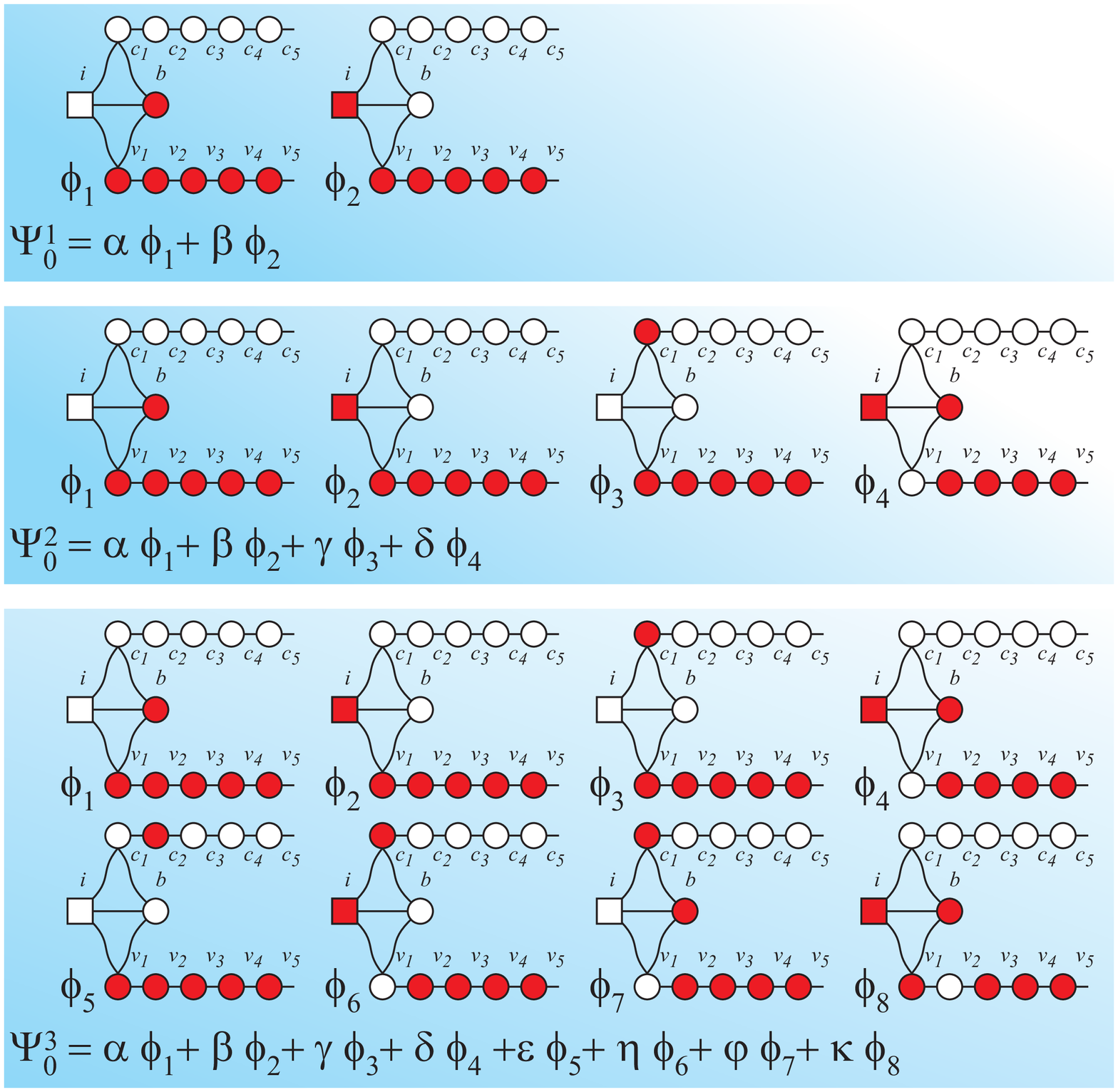}
    \caption{(color online) Graphical representation of the evolution of the basis set during the Lanczos cycles which determine the ground state wavefunction.}
    \label{Fig:Basis}
 \end{figure}

The number of Slater determinants available in the many particle basis is so large ($\approx 10^{100}$) that most of them have to be neglected. This is allowed as long as the total weight of the neglected states is small. In this section a method is discussed to find the $\approx 10^9$ determinants with the largest weight in a relatively short time period. The general physically relevant Hamiltonian is given in second quantization as:
\begin{equation}
\label{eq:Hlanc}
H=\sum_{\gamma,\gamma'} \epsilon_{\gamma,\gamma'} a^{\dag}_{\gamma}a^{\phantom{\dag}}_{\gamma'} + \sum_{\gamma,\gamma',\gamma'',\gamma'''} U_{\gamma,\gamma',\gamma'',\gamma'''} a^{\dag}_{\gamma}a^{\dag}_{\gamma'}a^{\phantom{dag}}_{\gamma''}a^{\phantom{\dag}}_{\gamma'''},
\end{equation}
with $\gamma$ an index for spin, orbital and site index (bath as well as impurity site) of the fermions included in the one-particle orbital basis. Note that the Hamiltonian in equation (\ref{eq:Hlanc}) is extremely general. The method described here to find the lowest $\approx 10^9$ determinants from a much larger basis set can be used for finite size lattice models with correlations (Heisenberg spin-exchange model, $tJ$ model, Hubbard model) \cite{LeTacon:2011ih, Glawion:2011jf}, ligand field theory calculations \cite{Haverkort:2012du}, or other forms of quantum chemistry models where one needs to diagonalize large sparse matrices.

The idea behind the method is to first define a relatively small basis, consisting of only a few Slater determinants, based on the Hartree-Fock or DFT energies of the orbitals. In this basis the ground state wave-function is found as a linear combination of the Slater determinants present in the basis. One then can rotate the one particle orbitals to minimize the number of Slater determinants needed as described in Appendix \ref{app:natorb}. If there are Slater determinants in the basis that do not contribute noticeably to the ground state wave function, these states are removed from the basis. Next the basis is enlarged by acting with the Hamiltonian on the ground state wave function in the small basis allowing all states that couple to this state but were not in the basis set to enter. One continues by finding the ground state wave-function in this new basis. These steps are repeated until convergence is reached, which can take up to a hundred loops. Nonetheless, finding the ground state even for rather involved basis sets is relatively fast (sub second on a laptop) as one starts with very small basis sets and each time the basis set is increased one can use the converged ground state calculation of the previous basis set as a starting point. In order to understand the basics of the algorithm, one can look at a graphical representation of the one-electron states or sites, single Slater determinant many electron basis states and multi Slater determinant eigenstates.

In figure \ref{Fig:Basis} the evolution of the basis states is shown. The one-electron states are represented by circles for the bath sites and a square for the impurity site. Solid circles are occupied, open circles are empty. The bath sites are labeled by $v_i$ for the valence bath, $c_i$ for the conduction bath and $b$ for one site at an energy such that its occupation is $1-n$ with $n$ the impurity occupation. In the top panel we show two basis functions, labeled $\phi_1$ and $\phi_2$. The valence bath sites are fully occupied (solid circles) and the conduction bath sites are completely empty (open circles). There is furthermore one-electron either at the impurity site $i$, or at the bath site labeled by $b$. This defines the two basis functions: $\phi_1$ and $\phi_2$. The ground state in this basis will be some linear combination of these two Slater determinants: $\psi_0^1 = \alpha \phi_1 + \beta \phi_2$.

Acting with the Hamiltonian on $\psi_0^1$ allows the electron from the valence bath site labeled $v_1$ to hop to either site $b$ or the impurity site, or allows the electron at the impurity site to hop to the conduction bath site $c_1$. $H \psi_0^1$ defines a new function $\varphi_0^2 = H \psi_0^1 / (\sqrt{\langle \psi_0^1 | H^2 | \psi_0^1 \rangle})$. In order to represent this new function one needs two more basis states as indicated in the middle panel of figure~\ref{Fig:Basis}: 
\begin{align}
\varphi_0^2 \propto & \langle \phi_1 | H | \psi_0^1 \rangle \phi_1 + \langle \phi_2 | H | \psi_0^1 \rangle \phi_2 \\
\nonumber           & + \langle \phi_3 | H | \psi_0^1 \rangle \phi_3 + \langle \phi_4 | H | \psi_0^1 \rangle \phi_4.
\end{align}
The basis states $\phi_1$ to $\phi_4$ span the new, larger basis. The function $\varphi_0^2$, in general, will not be an eigenstate in the new, larger basis. In this new basis, one can find, with the use of a Lanczos algorithm the new ground state without too much effort. The ground state in this basis will be, in general, some linear combination of four basis functions: $\psi_0^2 = \alpha \phi_1 + \beta \phi_2 + \gamma \phi_3 + \delta \phi_4$. Once the ground state in this new basis has been found, one can act with the full Hamiltonian on this state, which again will enlarge the basis needed to represent this new state. The third basis is shown in the bottom panel of figure~\ref{Fig:Basis}. The third ground state is given as some linear combination of these eight states.

Within this loop the size of the basis set grows exponentially and only a few steps can normally be done before the basis set size is so large that one cannot store the eigenstates any-more. The solution is to remove those basis states that do not noticeably contribute to the ground state. 

Given a basis set defined by the states $\phi_j$ and the ground state as $\psi_0 = \sum_j^{N_j} \alpha_j \phi_j $, all states $\phi_j$ are removed from the basis for which $\alpha_j^2<\epsilon$ with $\epsilon\approx 10^{-16}$. This new basis is then enlarged by acting with the Hamiltonian on the ground state ($\psi_0$) and adding those Slater determinants to the basis needed to represent $H \psi_0$. In this new basis the ground state is found, the determinants not needed to represent the ground state are removed and the basis is extended again by acting with the Hamiltonian on the ground state and adding those determinants needed to represent $H \psi_0$. This is repeated until convergence is reached, which can take up to 100 repetitions (generally less). For a converged calculation $\langle \psi_0 | H^2 | \psi_0 \rangle = \langle \psi_0 | H | \psi_0 \rangle^2$, which is fulfilled for all calculations in this paper up to the numerical accuracy ($\sim 10^{-14}$), one can obtain with floating point (double) precision. One should note that even for a converged calculation acting with the Hamiltonian on the ground state in a given basis will add states to the basis that were not included before. It is just that after diagonalization the new ground state has negligible weight in these determinants such that they are removed from the basis by the truncation procedure (and added again if one would go for another loop). It can also happen that states removed in an early loop of the calculation will reenter and become important in a later stage of the calculation.

The number of Slater determinants in the basis grows exponentially as a function of the number of steps in this algorithm. It is therefore of uttermost importance to remove those determinants that have a negligible contribution to the ground state. In order to find a ground state wave-function that has most of its weight in only a few determinants, one needs to optimize the one-particle orbitals. For a Hamiltonian where all states are correlated, this is the basis of natural orbitals. That is, in this case one rotates the one-particle orbitals after each calculation of the ground state such that the density matrix of the ground state is diagonal. For calculations on an impurity model this is not most efficient, as it mixes correlated impurity sites with noninteracting bath sites. The natural orbitals for an impurity model are discussed in Appendix~\ref{app:natorb}.

For a single band calculation the algorithm is rather straight-forward and robust. In a multiorbital case one needs to be slightly careful concerning the symmetry of the wave-function related to the starting point. For example, in the case of Co$^{3+}$ as found in LaCoO$_{3}$  one finds a local low spin state $t_{2g}^6$ ($S=0$) and local high spin state $t_{2g}^4 e_g^2$ ($S=2$) close in energy.\cite{Haverkort:2006hj} If one starts the algorithm from a low (high) spin initial state, one will (within the ligand field approximation) converge to the low (high) spin eigenstates.

\section{Optimizing the one-particle basis - Natural orbitals for impurity problems}
\label{app:natorb}

The DMFT equations as implemented in this work require one to calculate the ground state and Green's function of an Anderson impurity problem. Although only the impurity has correlations, an Anderson impurity model is still highly nontrivial and shows strong entanglement between the impurity and bath orbitals in the ground state. The ground state is generally not single-Slater-determinant representable. In order to minimize the number of Slater determinants needed to give a good representation of the ground state, we optimize the one-particle basis set. In this section we show how to do this.

We label the impurity site by $i$ and the bath sites by $b_j$, with $j\in [1,N_b]$. The impurity might have several internal degrees of freedom, as spin, orbital or site which will be labeled by a further quantum number $\tau$. The resulting Hamiltonian is:
\begin{align}
H_A =& \sum_{\tau,\tau',\tau'',\tau'''} U_{\tau,\tau',\tau'',\tau'''} a^{\dag}_{i,\tau}a^{\dag}_{i,\tau'}a^{\phantom{\dag}}_{i,\tau''}a^{\phantom{\dag}}_{i,\tau'''} \\
\nonumber&+ \sum_{\tau,\tau'} \alpha_{i,\tau;i,\tau'} a^{\dag}_{i,\tau}a^{\phantom{\dag}}_{i,\tau'} \\
\nonumber&+ \sum_{\tau,\tau'} \sum_j \beta_{i,\tau;b_j,\tau'} (a^{\dag}_{i,\tau}a^{\phantom{\dag}}_{b_j,\tau'} + a^{\dag}_{b_j,\tau'}a^{\phantom{\dag}}_{i,\tau}) \\
\nonumber&+ \sum_{\tau,\tau'} \sum_{j,j'} \alpha_{b_j,\tau;b_{j'},\tau'} a^{\dag}_{b_j,\tau}a^{\phantom{\dag}}_{b_{j'},\tau'}.
\end{align}
The aim is to find a unitary transformation of the one-particle states labeled by $\tau$, $i$, and $b_j$ such that the ground state can be represented by a minimum amount of Slater determinants. This transformation, however, should not mix impurity ($i$) states with bath states ($b_j$). If we label the transformed states by $t$, $\eta$, and $b_{\varsigma}$, we can define the unitary transformation $u$ such that
\begin{align}
a^{\dag}_{i,t} &= \sum_{\tau} u_{i,\tau;i,t} a^{\dag}_{i,\tau},\\\nonumber
a^{\dag}_{b_{\varsigma},t} &= \sum_{j,\tau} u_{j,\tau;\varsigma,t} a^{\dag}_{b_{j},\tau}.
\end{align} 
The transformation on the impurity $u_{i,\tau;i,t}$ is taken such that the density matrix of the ground state ($\psi_0$) of $H_A$ is diagonal:
\begin{equation}
n_{t,t'}^i = \langle \psi_0 | a^{\dag}_{i,t}a^{\phantom{\dag}}_{i,t'} | \psi_0 \rangle = \delta_{t,t'} n_{t,t'}^i.
\end{equation}
This is a trivial, noncostly step in the current method. The many-body ground state wave-function ($\psi_0$) is, as described in Appendix~\ref{app:lancsparse}, first calculated on a small basis, which is then gradually extended. After each calculation of the ground state for a given basis we calculate the density matrix of the impurity:
\begin{equation}
n_{\tau,\tau'}^i=\langle \psi_0 | a^{\dag}_{i,\tau}a^{\phantom{\dag}}_{i,\tau'} | \psi_0 \rangle.
\end{equation}
We can diagonalize this density matrix $n_{\tau,\tau'}^i$ and the eigenvectors of this matrix define the unitary transformation $u_{i,\tau;i,t}$. This transformation is applied to both $\psi_0$ and $H_A$. The loop to calculate the ground state is continued by extending the basis set as described in Appendix \ref{app:lancsparse}. 

The transformation of the bath states $u_{j,\tau;\varsigma,t}$ is less trivial. In principle, one would like to take the bath discretization to be defined such that the bath density matrix is diagonal in the basis chosen:
\begin{equation}
n_{\varsigma,t;\varsigma',t'}^b = \langle \psi_0 | a^{\dag}_{b_{\varsigma},t}a^{\phantom{\dag}}_{b_{\varsigma'},t'} | \psi_0 \rangle = \delta_{\varsigma,\varsigma'} \delta_{t,t'} n_{\varsigma,t;\varsigma',t'}^b.
\end{equation}
If for an arbitrary bath discretization one could calculate $\psi_0$, one can easily calculate the bath density matrix $a^{\dag}_{b_j,\tau}a^{\phantom{\dag}}_{b_{j'},\tau'}$ and the eigenvectors of this matrix define the optimal unitary transformation. The problem that arises though is that for an arbitrary bath discretization all bath states are important and one cannot truncate the many-body wavefunction such that only a few (maximally $\approx 10^9$) Slater determinants are needed to represent the wavefunction. Once a solution is found, we can define a basis that would have been more efficient, but we need to define the efficient basis before the calculation can be done. The iterative method, which works well for the impurity sites, is impractical for the bath sites as the number of orbitals involved is too large.

We need to define a unitary transformation $u_{j,\tau;\varsigma,t}$ that approximately leads to a diagonal density matrix, but can be calculated before the many-body problem is solved. This is done by introducing a noninteracting reference system: 
\begin{align}
\widetilde{H}_A =& \sum_{\tau,\tau'} V_{\tau,\tau'} a^{\dag}_{i,\tau}a^{\phantom{\dag}}_{i,\tau'} \\
\nonumber&+ \sum_{\tau,\tau'} \alpha_{i,\tau;i,\tau'} a^{\dag}_{i,\tau}a^{\phantom{\dag}}_{i,\tau'} \\
\nonumber&+ \sum_{\tau,\tau'} \sum_j \beta_{i,\tau;b_j,\tau'} (a^{\dag}_{i,\tau}a^{\phantom{\dag}}_{b_j,\tau'} + a^{\dag}_{b_j,\tau'}a^{\phantom{\dag}}_{i,\tau}) \\
\nonumber&+ \sum_{\tau,\tau'} \sum_{j,j'} \alpha_{b_j,\tau;b_{j'},\tau'} a^{\dag}_{b_j,\tau}a^{\phantom{\dag}}_{b_{j'},\tau'}.
\end{align}
$V$ is chosen such that the correct impurity occupation is reproduced. The correct impurity occupation is known from a previous step in the calculation, which is either a previous DMFT loop or a previous calculation with a smaller many-body basis set. The solution of the reference system, which only has one body interactions, is trivial. Diagonalization of $\widetilde{H}_A$ leads to a set of one particle states that are a mixture of bath ($b_j,\tau$) and impurity ($i,\tau$) states. The many-body ground state ($\psi_0$) is a single Slater determinant in which all one particle eigenstates of $\widetilde{H}_A$ with an energy smaller than the chemical potential are occupied. From this state we can calculate the bath density matrix:
\begin{equation}
n_{j,\tau;j';\tau'}^b=\langle \psi_0 | a^{\dag}_{b_j,\tau}a^{\phantom{\dag}}_{b_{j'},\tau'} | \psi_0 \rangle.
\end{equation}
We can diagonalize this density matrix $n_{j,\tau;j',\tau'}^b$ and the eigenvectors of this matrix define the unitary transformation $u_{j,\tau;\varsigma,t}$. This transformation is applied to the bath and thereby minimizes the number of Slater determinants needed in the calculation.

The optimized bath which leads to a diagonal density matrix ($n_{\varsigma,t;\varsigma',t'}^b$) for the ground state of the reference system ($\widetilde{H}_A$) always has the same form. The resulting bath geometry is depicted in figure \ref{Fig:Structure} (c). It is interesting to study this bath geometry in a bit more detail. For the reference system the impurity will have an occupation $n$, the bath site labeled by $b$ has an occupation $1-n$. The bath sites labeled by $v_j$ are all fully occupied and the bath sites labeled by $c_j$ are all completely empty. The many-body ground state for the reference system is given by only four Slater determinants with only partially filled states $i$ and $b$, which define a molecular bond between these two states. For a single band impurity with $\tau$ labeling spin up ($\uparrow$) and spin down ($\downarrow$) states, this function can be written as:
\begin{equation}
|\psi_0\rangle = (\alpha a^{\dag}_{i,\uparrow} + \beta a^{\dag}_{b,\uparrow})(\alpha a^{\dag}_{i,\downarrow} + \beta a^{\dag}_{b,\downarrow})\Pi_{j=1}^{j=N_v} a^{\dag}_{v_j,\uparrow}a^{\dag}_{v_j,\downarrow}|0\rangle,
\end{equation}
with $\alpha$ and $\beta$ positive parameters such that $\alpha^2+\beta^2=1$ and the indices as shown in figure \ref{Fig:Structure}(c). 

The bath sites labeled by $v_j$ are fully occupied in the ground state and the bath sites labeled by $c_j$ are completely empty. Nonetheless, there is an interaction between the impurity site and these bath sites. If the interaction between sites $i$ and $c_1$ ($v_1$) is $t_{ic}$ ($t_{iv}$) and the interaction between sites $b$ and $c_1$ ($v_1$) is given by $t_{bc}$ ($t_{bv}$), respectively, then the relation between these interactions is
\begin{align}
\nonumber
\alpha t_{ic} + \beta t_{bc} &= 0,\\
-\beta t_{iv} + \alpha t_{bv} &= 0.
\end{align}
The interaction between the occupied states at site $i$ with the unoccupied conduction bath sites $c_1$ interferes with the hopping from the occupied state at $b$ such that the total interaction cancels. 

The basis obtained in the reference system is used as the basis for the correlated Anderson impurity problem. Here the molecular orbital formed between the states $i$ and $b$ becomes partly unoccupied as one moves towards the Heitler-London solution for correlated molecular bonds. The choice of this basis allows one to select a few Slater determinants that are important. For the calculations presented in this paper we never needed more than a few thousand determinants to represent the ground state.

\section{Reduction of the number of poles}
\label{app:polered}

The number of poles in the bath Green's function defines the number of bath sites in the Anderson impurity Hamiltonian. The current algorithm is able to include several hundreds of such states. The new bath Green's function has a dimension of $N_{b}^{new} = N_0 \times (N_{\Sigma} + 1)$, equation (\ref{eq:bathnew}) and $N_{\Sigma} = N_c$, equation (\ref{eq:sigma}). The number of poles in the bath Green's function thus grows rapidly with each self-consistency loop and needs to be reduced. Following the ideas of renormalization group theory, one could choose a fixed set of energies on a logarithmic mesh that is used to represent the bath Green's function. Although not a bad choice, especially as it allows one to represent the Fermi energy with a large number of poles, we here opt for an adaptive mesh. We want the Green's function to be represented by a large number of poles in those areas where the Green's function is large and by a smaller number of poles where the Green's function is small. In practice, we repeatedly remove the pole with the smallest spectral weight and merge this pole with the neighboring poles until the number of poles is reduced to the number of bath orbitals one wants to include in the calculation. The same procedure is used to remove poles with a negative weight from the self-energy.

Starting from a Green's function or self-energy represented as
\begin{equation}
G(\omega)=\sum_{j=1}^{N} \frac{{\beta}_{j}^2}{\omega-{\alpha}_{j}},
\end{equation}
with $\alpha_{j}<\alpha_{j+1}$. We repeatedly determine the pole with the smallest weight (minimal $\beta_{j}^2$) and remove this pole from the Green's function, whereby we locally keep the zeroth and first moments conserved. Assuming that the pole with smallest weight is found at position $k$, then after one iteration this leads to the Green's function,
\begin{equation}
G'(\omega)=\sum_{j=1}^{N-1} \frac{{\beta'}_{j}^2}{\omega-{\alpha'}_{j}},
\end{equation}
with ${\alpha'}_{j}=\alpha_{j}$ (${\beta'}_{j}^2=\beta_{j}^2$) for all $j \leq k-2$, ${\alpha'}_{j}={\alpha}_{j+1}$ (${\beta'}_{j}^2={\beta}_{j+1}^2$) for all $k+1 \leq j \leq N-1$, and:
\begin{eqnarray}
          {\beta'}_{k-1}^2 &=& \beta_{k-1}^2 + \frac{\alpha_{k+1}-\alpha_{k}}{\alpha_{k+1}-\alpha_{k-1}} \beta_{k}^2,\\
\nonumber {\beta'}_{k}^2   &=& \beta_{k+1}^2 + \frac{\alpha_{k}-\alpha_{k-1}}{\alpha_{k+1}-\alpha_{k-1}} \beta_{k}^2,\\
\nonumber {\alpha'}_{k-1}  &=& \frac{\alpha_{k-1} \beta_{k-1}^2 (\alpha_{k+1}-\alpha_{k-1})+\alpha_{k} \beta_{k}^2 (\alpha_{k+1}-\alpha_{k})}{ \phantom{\alpha_{k-1}} \beta_{k-1}^2 (\alpha_{k+1}-\alpha_{k-1})+\phantom{\alpha_{k}} \beta_{k}^2 (\alpha_{k+1}-\alpha_{k}) },\\
\nonumber {\alpha'}_{k}    &=& \frac{\alpha_{k+1} \beta_{k+1}^2 (\alpha_{k+1}-\alpha_{k-1})+\alpha_{k} \beta_{k}^2 (\alpha_{k}-\alpha_{k-1})}{\phantom{\alpha_{k+1}} \beta_{k+1}^2 (\alpha_{k+1}-\alpha_{k-1})+\phantom{\alpha_{k}} \beta_{k}^2 (\alpha_{k}-\alpha_{k-1})}.
\end{eqnarray}

The equations look more involved than they are. The weight of the pole at positions $k$ is split between the weight at position $k-1$ and $k+1$ weighted by the distance to the neighboring pole. The energy ($\alpha_{k-1}$) of pole $k-1$ is shifted such that the first moment of poles $k-1$ and $k$ (multiplied by the partial weight) is conserved (the same is true for $\alpha_{k+1}$.):
\begin{eqnarray}
{\beta'}_{k-1}^2 {\alpha'}_{k-1} &=& \beta_{k-1}^2 \alpha_{k-1} +  \alpha_{k} \beta_{k}^2 \frac{\alpha_{k+1}-\alpha_{k}}{\alpha_{k+1}-\alpha_{k-1}},\\
\nonumber {\beta'}_{k}^2 {\alpha'}_{k} &=& \beta_{k+1}^2 \alpha_{k+1} +  \alpha_{k} \beta_{k}^2 \frac{\alpha_{k}-\alpha_{k-1}}{\alpha_{k+1}-\alpha_{k-1}}.
\end{eqnarray} 
This procedure conserves locally the zeroth and first moment,
\begin{eqnarray}
{\beta'}_{k-1}^2 + {\beta'}_{k}^2  &=& \beta_{k-1}^2 + \beta_{k}^2 + \beta_{k+1}^2,\\
\nonumber {\beta'}_{k-1}^2 {\alpha'}_{k-1} + {\beta'}_{k}^2 {\alpha'}_{k} &=& \beta_{k-1}^2 \alpha_{k-1} + \beta_{k}^2 \alpha_{k} +\beta_{k+1}^2 \alpha_{k+1},
\end{eqnarray}
and nonlocally only introduces small errors in the moments of the full Green's function. 

It should be noted that the shift in spectral weight introduced by this procedure is generally small. For a spectrum with bandwidth $W$ represented by $N$ poles the maximum shift of a single pole (spectral weight times distance) is of order $W/N^2$ and thus converges well as a function of $N$. It is expected that many different algorithms to reduce the number of poles will yield similar results and the procedure presented here might not be the optimum. Any algorithm used should reduce the poles at large energy ($|\alpha|\gg W$) with very small weight ($\beta^2 \ll 10^{-7}$) as these are most probably spurious eigenstates introduced by the Lanczos algorithm and destabilize the self-consistency loops.

\end{document}